\documentclass{article}
\pdfoutput=1

\usepackage[preprint,nonatbib]{neurips_2024} % nonatbib to disable natbib for simplicity
\usepackage[numbers]{natbib} % Load natbib with numerical citations

\usepackage[utf8]{inputenc} % allow utf-8 input
\usepackage[T1]{fontenc}    % use 8-bit T1 fonts
\usepackage{hyperref}       % hyperlinks
\usepackage{url}            % simple URL typesetting
\usepackage{booktabs}       % professional-quality tables
\usepackage{amsfonts}       % blackboard math symbols
\usepackage{nicefrac}       % compact symbols for 1/2, etc.
\usepackage{microtype}      % microtypography
\usepackage{verbatim}  % For creating verbatim text sections
\usepackage{enumitem}  % Enhanced list environments
\usepackage{graphicx}
\usepackage{xcolor}
\usepackage{longtable}

\title{Fair Knowledge Tracing in Second Language Acquisition}

\author{%
  Weitao Tang \\
  Monash University \\
  \texttt{Weitao.Tang1@monash.edu} \\
  \And
  Dr. Guanliang Chen\thanks{Corresponding author: \texttt{guanliang.chen@monash.edu}} \\
  Monash University \\
  \texttt{guanliang.chen@monash.edu} \\
  \And
  Shuaishuai Zu \\
  Renmin University of China \\
  \texttt{Zushuaishuai@gmail.com} \\
  \And
  Jiangyi Luo \\
  Australian National University \\
  \texttt{Jiangyi.Luo@anu.edu.au}
}

\begin{document}

\maketitle

\begin{abstract}
In the domain of second-language acquisition, predictive modeling serves as a pivotal tool for facilitating educators in implementing diversified teaching strategies, thereby garnering extensive research attention. Despite the prevalent focus on model accuracy in most existing studies, the exploration into model fairness remains substantially underexplored. Model fairness pertains to the equitable treatment of different groups by machine learning algorithms. It ensures that the model's predictions do not exhibit unintentional biases against certain groups based on attributes such as gender, ethnicity, age, or other potentially sensitive characteristics. In essence, a fair model should produce outcomes that are impartial and do not perpetuate existing prejudices, ensuring that no group is systematically disadvantaged. In this research, we evaluate the fairness of two predictive models based on second-language learning, utilizing three tracks from the Duolingo dataset: en\_es (English learners who speak
Spanish), es\_en(Spanish learners who speak English),
and fr\_en(French learners who speak English). We measure 
\begin{enumerate}[label=(\roman*)]
    \item algorithmic fairness among different clients such as iOS, Android and Web and 
     \item algorithmic fairness between developed countries and developing countries.
\end{enumerate}
Our findings indicate: 
\begin{enumerate}[label=\arabic*)]
    \item Deep learning exhibits a marked advantage over machine learning when applied to knowledge tracing based on second language acquisition, owing to its heightened accuracy and fairness.  
    \item Both machine learning and deep learning algorithms exhibit a noticeable bias favoring mobile users over their non-mobile counterparts.
    \item Compared to deep learning algorithms, machine learning algorithms showcase a more pronounced bias against developing countries.
    \item To strike a balanced approach in terms of fairness and accuracy, deep learning is identified as being more apt for making predictions in the \texttt{en\_es} and \texttt{es\_en} tracks, whereas machine learning emerges as a more suitable option for the \texttt{fr\_en}.
\end{enumerate}
This study underscores the necessity to delve deeper into the realms of fairness in predictive models, ensuring equitable educational strategies across diverse clients and countries.
\end{abstract}
\newpage

\section{Introduction}

The historical reliance on traditional metrics like student performance, feedback, and participation to shape educational methodologies was fraught with potential inaccuracies [\citenum{price2010feedback}]. Such inaccuracies mainly arise because humans cannot process the vast amount of data generated during the educational process like machines can. A deeper delve suggests that these inaccuracies might be influenced by a range of psychological factors [\citenum{culverhouse2003experts}]. For instance, educators might be influenced by \begin{enumerate}
    \item The limits of short-term memory.
    \item Feelings of fatigue or boredom.
    \item A positivity bias, where there is a predisposition towards favoring positive feedback from students.(This means that an educator might harbor an undue expectation for positive comments, thereby skewing their interpretation of student feedback)
\end{enumerate}
 As a result, a single poor examination outcome might prompt an educator to simplify the test content, risking a dilution of educational standards. To mitigate these issues, in 1994, Corbett and Anderson introduced the concept of "knowledge tracing" [\citenum{corbett1994knowledge}]. The crux of this approach was to harness computational technology to model various student indicators such as grades, feedback, participation, and other features. By doing so, predictions about students' future performance could be made, enabling the crafting of more appropriate educational strategies.

Employing computational techniques effectively reduces the chances of human-induced errors, enhancing the precision of outcomes. In recent times, educational technology increasingly incorporates artificial intelligence to employ data and predictive models [\citenum{holstein2017intelligent,pedro2019artificial,roll2016evolution,luckin2016intelligence,bienkowski2012enhancing}]. This provides tailored support and insights for students, educators, and administrators alike [\citenum{baker2014educational,luckin2019designing}]. This process closely resembles an adaptive system. As students accumulate knowledge, the system makes predictions regarding their understanding and offers varied learning materials tailored to their needs [\citenum{pane2010experiment}].

\begin{figure}[h]
  \centering
  \includegraphics[width=10cm,height=6cm]{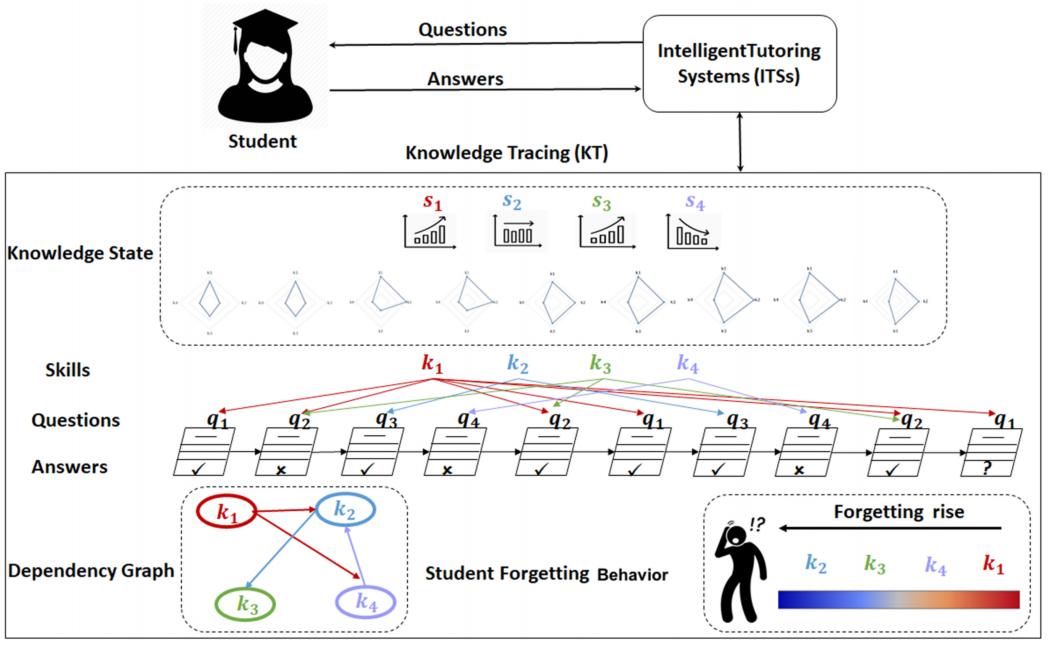}
  \caption{Intelligent tutoring system [\citenum{Abdelrahman2023}].}
  \label{fig:IntelligentTutoringSystem}
\end{figure}

As outlined by [\citenum{Abdelrahman2023}], the Intelligent Tutoring System (ITS Figure \ref{fig:IntelligentTutoringSystem}) functions as an adaptive feedback mechanism. The ITS poses questions to students and, based on their responses, discerns their knowledge state. This knowledge state offers insight into the student's proficiency in various skills. For instance, if a student consistently demonstrates over 90\% accuracy in addition tasks, one could infer mastery over the skill of addition. However, the same student might struggle with problems combining addition and subtraction, indicating a lack of proficiency in the latter. Certain skills might be interdependent; for example, mastering calculus presupposes proficiency in basic arithmetic operations like addition and subtraction. As learning progresses, students might forget certain concepts, with more complex skills generally having a higher forgetting rate.

While contemporary knowledge tracing algorithms have achieved commendable accuracy, there remains a pressing need to explore their fairness. For instance, gender discrepancies could lead to algorithmic biases [\citenum{datta2014automated,simonite2015probing,duran2020gender,rayyan2013participation}]. Similar biases, such as racial prejudices, have been observed in criminal justice record systems [\citenum{angwin2022machine}]. As concerns over fairness in predictive educational models grow [\citenum{gardner2019evaluating}], ensuring algorithmic equity becomes pivotal for creating an inclusive educational environment. For diverse student populations, investigating the influence of factors like native language background on algorithmic fairness is crucial.

Second language acquisition is a vital area of study, shaping our understanding of cognitive processes, cultural exchange, and global communication. The ability to acquire a second language not only bridges cultural and geographical divides but also offers individuals a competitive edge in today's globalized world. It fosters cognitive benefits, including problem-solving skills, creativity, and memory enhancements. With the increasing reliance on technology to aid language learning, ensuring the fairness and effectiveness of these tools is paramount.

The motivation behind this research arises from the pressing need to ensure equity in technological tools employed for second language acquisition. While machine learning and deep learning models promise unprecedented benefits in tailoring learning experiences, there's a pressing need to scrutinize potential biases they may harbor. The objective of this study is to delve into fairness concerns in machine learning and deep learning models using the Duolingo dataset. Specifically, we intend to utilize the Gradient Boosted Decision Trees (GBDT) algorithm and the Multi-task Learning algorithm—both of which have demonstrated remarkable results in the Duolingo shared task\footnote{\href{https://dataverse.harvard.edu/dataset.xhtml?persistentId=doi:10.7910/DVN/8SWHNO}{Data for the 2018 Duolingo Shared Task on Second Language Acquisition Modeling (SLAM)}}. Our primary goal is to ascertain the extent to which these algorithms exhibit algorithmic biases. We investigate biases concerning students' demographic attributes, particularly "country" and "client." Here, "country" is categorized into developed and developing nations, highlighting socio-economic disparities. Simultaneously, "client" delineates the platforms students use for learning, encompassing iOS, Android, and Web interfaces. By understanding these biases, this research endeavors to offer insights that can direct developers in forging more equitable language learning tools and platforms.

\section{Related Work}
In the scope of this project, a compilation of 16 research articles was undertaken, all of which explore the application of knowledge tracing on the Duolingo dataset. These research contributions can be broadly bifurcated into three primary categories: those employing Traditional Machine Learning-based Algorithms (7 articles), and those utilizing Deep Learning-based Algorithms (9 articles), with the objective of predicting student performance metrics. A notable 13 of these papers presented predictions across all three tracks, while a pair exclusively targeted the English track, and a singular article was dedicated to investigating both the Spanish and French tracks. Subsequent sections will delve into a meticulous analysis of the models adopted in these articles, along with an exploration of their respective performance metrics in a detailed manner.

\subsection{Methodological Approaches}
To better compare the models in these 16 papers, two metrics will be introduced. One is the F1 score, which represents the harmonic mean of Precision and Recall. The other is AUC (Area Under the Curve) [\citenum{bradley1997use}]. AUC is the area under the ROC (Receiver Operating Characteristic) curve, which primarily describes the relationship between the False Positive Rate and the True Positive Rate. Both values are closer to 1, indicating a better model.

\subsection{Comparison between all Algorithms}
Table~\ref{tab:all} illustrates that models integrating deep learning or combining it with machine learning tend to outperform, whereas standalone machine learning models might fall short in certain tasks. This points towards a necessity for neural networks in Second Language Acquisition (SLA) tasks to delve deeper into feature extraction, considering the Duolingo dataset may present complexities beyond the capacity of traditional machine learning models. Notably, [\citenum{hu2020multi}] delivered superior outcomes by employing an encoder-decoder architecture and multi-task learning, which facilitated the use of shared representation layers and the extraction of features even in data-scarce scenarios. Conversely, the LM-KT model presented in [\citenum{srivastava2021question}] exhibited subpar performance, potentially due to intrinsic model constraints and a lack of comprehensive model optimization by the authors. Hence, the selection of an appropriate model emerges as pivotal in the tracing of knowledge in SLA. Both high-achieving and underachieving models have vitally informed the evolution of SLA knowledge tracing. In summary, within the realm of Machine Learning (ML) algorithms, the most notable results were achieved by [\citenum{rich2018modeling}], while [\citenum{hu2020multi}] claimed the pinnacle of performance among Deep Learning (DL) algorithms, thanks to its utilization of multi-task learning.

\begin{table}[ht]
    \centering
    \caption{Comparison of all model's performance}
    \label{tab:all}
    
    \begin{minipage}{.3\textwidth}
        \centering
        \textbf{(a) English Track}
        \small
        \begin{tabular}{lcc}
            \toprule
            Literature & AUC$\downarrow$ & F1$\downarrow$ \\
            \midrule
            {[\citenum{hu2020multi}]$\bullet$} & .864 & .564 \\
           {[\citenum{ruan2021variational}]$\bullet$} & .863 & .564 \\
           {[\citenum{osika2018second}]$\bullet$} & .861 & .561 \\
           {[\citenum{xu2018cluf}]$\bullet$} & .861 & .559 \\
           {[\citenum{rich2018modeling}]$\bigtriangleup$} & .859 & .468 \\
           {[\citenum{sense2022cognition}]$\bigtriangleup$} & .854 &  -\\
           {[\citenum{palenzuela2022modeling}]$\bullet$} & .853 & .479 \\
           {[\citenum{kaneko2018tmu}]$\bullet$} & .848 & .476 \\
           {[\citenum{bestgen2018predicting}]$\bigtriangleup$} & .846 & .414 \\
           {[\citenum{yuan2018neural}]$\bullet$} & .841 & .479 \\ 
           {[\citenum{tomoschuk2018memory}]$\bigtriangleup$} & .829 & .424 \\
           {[\citenum{chen2018feature}]$\bigtriangleup$} & .821 & .389 \\
           {[\citenum{nayak2018context}]$\bigtriangleup$} & .821 & .376 \\
           {[\citenum{klerke2018grotoco}]$\bigtriangleup$} & .817 & .462 \\
           {[\citenum{vie2018deep}]$\bullet$} & .815 & .329 \\
           {[\citenum{srivastava2021question}]$\bullet$}  & - & - \\
            \bottomrule
        \end{tabular}
    \end{minipage}
    \hspace{10pt}
    \begin{minipage}{.3\textwidth}
        \centering
        \textbf{(b) Spanish Track}
        \small
        \begin{tabular}{lcc}
            \toprule
            Literature & AUC$\downarrow$ & F1$\downarrow$ \\
            \midrule
            {[\citenum{hu2020multi}]$\bullet$} & .839 & .530 \\
            {[\citenum{ruan2021variational}]$\bullet$} & .838 & .531 \\
            {[\citenum{osika2018second}]$\bullet$} & .838 & .530 \\
            {[\citenum{xu2018cluf}]$\bullet$} & .835 & .524 \\
            {[\citenum{rich2018modeling}]$\bigtriangleup$} & .835 & .420 \\
            {[\citenum{kaneko2018tmu}]$\bullet$} & .824 & .439 \\
            {[\citenum{bestgen2018predicting}]$\bigtriangleup$} & .818 & .390 \\
            {[\citenum{yuan2018neural}]$\bullet$} & .807 & .435 \\
            {[\citenum{tomoschuk2018memory}]$\bigtriangleup$} & .803 & .375 \\
            {[\citenum{chen2018feature}]$\bigtriangleup$} & .801 & .344 \\
            {[\citenum{klerke2018grotoco}]$\bigtriangleup$} & .791 & .452 \\
            {[\citenum{nayak2018context}]$\bigtriangleup$} & .790 & .338 \\
            {[\citenum{vie2018deep}]$\bullet$} & .788 & .306 \\
            {[\citenum{srivastava2021question}]$\bullet$} & .750 & - \\
            {[\citenum{palenzuela2022modeling}]$\bullet$} & - & - \\
            {[\citenum{sense2022cognition}]$\bigtriangleup$} & - & - \\
            \bottomrule
        \end{tabular}
    \end{minipage}
    
    \bigskip  % Add some vertical space between the rows of subtables
    
    \begin{minipage}{.3\textwidth}
        \centering
        \textbf{(c) French Track}
        \small
        \begin{tabular}{lcc}
            \toprule
            Literature & AUC$\downarrow$ & F1$\downarrow$ \\
            \midrule
            {[\citenum{hu2020multi}]$\bullet$} & .860 & .579 \\
            {[\citenum{ruan2021variational}]$\bullet$} & .859 & .575 \\
            {[\citenum{osika2018second}]$\bullet$} & .857 & .573 \\
            {[\citenum{xu2018cluf}]$\bullet$} & .854 & .569 \\
            {[\citenum{rich2018modeling}]$\bigtriangleup$} & .854 & .493 \\
            {[\citenum{bestgen2018predicting}]$\bigtriangleup$} & .843 & .487 \\
            {[\citenum{kaneko2018tmu}]$\bullet$} & .839 & .502 \\
            {[\citenum{yuan2018neural}]$\bullet$} & .835 & .508 \\
            {[\citenum{tomoschuk2018memory}]$\bigtriangleup$} & .823 & .442 \\
            {[\citenum{chen2018feature}]$\bigtriangleup$} & .815 & .415 \\
            {[\citenum{klerke2018grotoco}]$\bigtriangleup$} & .813 & .502 \\
            {[\citenum{nayak2018context}]$\bigtriangleup$} & .811 & .431 \\
            {[\citenum{vie2018deep}]$\bullet$} & .809 & .406 \\
            {[\citenum{srivastava2021question}]$\bullet$} & .730 & - \\
            {[\citenum{palenzuela2022modeling}]$\bullet$} & - & - \\
            {[\citenum{sense2022cognition}]$\bigtriangleup$} & - & - \\
            \bottomrule
        \end{tabular}
    \end{minipage}
    
    \bigskip  % Add some vertical space between the subtables and the note
    
    \centering
    {$\bigtriangleup$} means using Machine Learning Algorithm, {$\bullet$} means using Deep Learning Algorithm
\end{table}

\section{Experiment}
\subsection{Experimental Environment and Set Up}
All experiments were conducted on a high-performance Alienware X17R2 laptop. The specifications of the device are as follows:

\textbf{Device:} Alienware X17R2 Laptop \\
\textbf{CPU:} Intel i9-12900HK \\
\textbf{Memory:} 64GB DDR5

The experiments utilized TensorFlow for both training the models on the CPU. 

To fully explore the fairness of algorithms on the Duolingo dataset, we selected the best-performing GBDT algorithm in machine learning [\citenum{rich2018modeling}] and the best-performing Multi-task learning in deep learning [\citenum{hu2020multi}] from Table~\ref{tab:all}. Based on this, we further explore the accuracy and fairness of these two algorithms.

\textbf{Accuracy metrics}
Consistent with previous knowledge tracing modeling based on second language acquisition, we employed two metrics to gauge the predictive accuracy of knowledge tracing: Area Under the Curve (AUC) and F1 score.

\textbf{Fairness metrics}
The study [\citenum{gardner2019evaluating}] is noted to be the inaugural exploration into identifying apt metrics for evaluating the fairness of predictive models in educational research. Specifically, a metric named Absolute Between-group ROC Area (ABROCA) was introduced, aiming to measure the predictive unfairness across various demographic groups by calculating the definite integral between the ROC curves of two observed groups. Significantly, ABROCA possesses two advantages:

\begin{enumerate}
    \item It accounts for performance disparities across the entire threshold spectrum, superseding other fixed-threshold approaches.
    \item It can be easily calculated from predictive results without the need for additional data or metrics.
\end{enumerate}

Thus, this metric was utilized in our research. It's crucial to note that a lower ABROCA value signifies reduced algorithmic unfairness in the predictive model.
We executed training using GBDT and multi-task learning across three tracks: \texttt{en\_es}, \texttt{es\_en}, and \texttt{fr\_en}. The data post-training were categorized in two dimensions:

\begin{enumerate}
    \item By countries, segmented into:
    \begin{itemize}
        \item Developed Countries
        \item Developing Countries
    \end{itemize}
    \item By client, segmented into:
    \begin{itemize}
        \item Android
        \item iOS
        \item Web
    \end{itemize}
\end{enumerate}

\vspace{1em} % Adds some space between the two paragraphs

Testing was subsequently conducted on the three tracks, employing AUC and F1 score as metrics to evaluate precision for developed and developing countries, and Android, iOS, and Web clients respectively. Concurrently, ABROCA was utilized to assess fairness.

\subsection{Accuracy}

\subsubsection{Accuracy on three tracks (en\_es, es\_en, fr\_en)}

\begin{table}[h]
    \centering
    \caption{Accuracy on Different Tracks}
    \begin{tabular}{llcc}
        \toprule
        Track & Model             & F1 Score & AUC \\
        \midrule
        EN\_ES & Multi-task learning & 0.554   & 0.858 \\
               & GBDT                & 0.469   & 0.856 \\
        ES\_EN & Multi-task learning & 0.519   & 0.830  \\
               & GBDT                & 0.421   & 0.833 \\
        FR\_EN & Multi-task learning & 0.571   & 0.855 \\
               & GBDT                & 0.493   & 0.851 \\
        \bottomrule
    \end{tabular}
\end{table}
\begin{enumerate}

\newpage
\item \textbf{Dataset Size Consideration}:
   \begin{itemize}
       \item The \textit{fr\_en} track corresponds to a smaller dataset compared to \textit{en\_es} and \textit{es\_en}. The complex nature of the Multi-task learning model may not be as beneficial for smaller datasets, resulting in suboptimal performance.
   \end{itemize}

\item \textbf{Training Time}:
   \begin{itemize}
       \item The Multi-task learning model, trained using TensorFlow-CPU, took approximately 11 days, making it a time-intensive approach. In contrast, the GBDT model took just an hour. This stark difference in training time suggests that while Multi-task learning may offer performance benefits in some scenarios, it comes at the cost of significantly increased computational time.
   \end{itemize}

\item \textbf{Performance Across Tracks}:
   \begin{itemize}
       \item \textit{EN\_ES Track}: The Multi-task learning model outperformed GBDT in F1 score for this track. However, their AUCs are very similar, suggesting both models are comparable in terms of ranking the predictions.
       \item \textit{ES\_EN Track}: While the Multi-task learning model still leads in F1 score, the gap is narrower. Both models' AUCs remain close.
       \item \textit{FR\_EN Track}: Multi-task learning exceeds GBDT in F1 score. However, considering the significantly shorter training time of GBDT and its competitive performance, GBDT might be a more feasible choice, especially for the smaller \textit{FR\_EN} dataset.
   \end{itemize}

\end{enumerate}
\subsubsection{Accuracy on Client (ios, android, web)}

\begin{table}[h]
    \centering
    \caption{Accuracy on Different Clients}
    \begin{tabular}{llcc}
        \toprule
        Track                & Model             & F1 Score & AUC \\
        \midrule
        CLIENT\_IOS\_EN\_ES    & Multi-task learning & 0.535   & 0.86  \\
                              & GBDT                & 0.434   & 0.86  \\
        CLIENT\_ANDROID\_EN\_ES & Multi-task learning & 0.551   & 0.859 \\
                              & GBDT                & 0.468   & 0.858 \\
        CLIENT\_WEB\_EN\_ES     & Multi-task learning & 0.584   & 0.849 \\
                              & GBDT                & 0.507   & 0.842 \\
        CLIENT\_IOS\_ES\_EN     & Multi-task learning & 0.528   & 0.831 \\
                              & GBDT                & 0.431   & 0.833 \\
        CLIENT\_ANDROID\_ES\_EN  & Multi-task learning & 0.524   & 0.84  \\
                              & GBDT                & 0.427   & 0.841 \\
        CLIENT\_WEB\_ES\_EN      & Multi-task learning & 0.475   & 0.8   \\
                              & GBDT                & 0.371   & 0.811 \\
        CLIENT\_IOS\_FR\_EN      & Multi-task learning & 0.421   & 0.727 \\
                              & GBDT                & 0.498   & 0.858 \\
        CLIENT\_ANDROID\_FR\_EN   & Multi-task learning & 0.4     & 0.719 \\
                              & GBDT                & 0.474   & 0.857 \\
        CLIENT\_WEB\_FR\_EN       & Multi-task learning & 0.44    & 0.713 \\
                              & GBDT                & 0.504   & 0.824 \\
        \bottomrule
    \end{tabular}
\end{table}

\begin{itemize}
    \item \textbf{Performance Across Clients}:
    \begin{itemize}
        \item \textit{iOS}: Multi-task learning consistently shows a higher F1 score across all tracks for iOS users. This might indicate that iOS users, who are often associated with higher spending power, benefit more from the complexity of the Multi-task learning model.
        \item \textit{Android}: The performance between the models is closer on this platform. For some tracks, Multi-task learning has a slight edge in F1 score.
        \item \textit{Web}: The F1 score of the Multi-task learning model is generally higher than that of GBDT, but GBDT sometimes leads in AUC, particularly for the \textit{FR\_EN} track.
    \end{itemize}
\end{itemize}

\newpage
\subsubsection{Accuracy on Country (developed countries and developing countries)}

\begin{table}[htbp]
    \centering
    \caption{Accuracy on Different Countries}
    \begin{tabular}{llcc}
        \toprule
        Track                & Model             & F1 Score & AUC \\
        \midrule
        COUNTRY\_DEVELOPED\_EN\_ES    & Multi-task learning & 0.553 & 0.863 \\
                              & GBDT                & 0.476 & 0.865 \\
        COUNTRY\_DEVELOPING\_EN\_ES & Multi-task learning & 0.555 & 0.856 \\
                              & GBDT                & 0.467 & 0.853 \\
        COUNTRY\_DEVELOPED\_ES\_EN     & Multi-task learning & 0.525 & 0.832 \\
                              & GBDT                & 0.428 & 0.835 \\
        COUNTRY\_DEVELOPING\_ES\_EN  & Multi-task learning & 0.480 & 0.818 \\
                              & GBDT                & 0.368 & 0.820 \\
        COUNTRY\_DEVELOPED\_FR\_EN      & Multi-task learning & 0.427 & 0.726 \\
                              & GBDT                & 0.497 & 0.851 \\
        COUNTRY\_DEVELOPING\_FR\_EN   & Multi-task learning & 0.410 & 0.729 \\
                              & GBDT                & 0.491 & 0.854 \\
        \bottomrule
    \end{tabular}
\end{table}

\begin{itemize}
    \item \textbf{Performance Across Countries}:
    \begin{itemize}
        \item \textit{Developed Countries}: The performance of both models is closely matched in developed countries. Interestingly, for the \textit{FR\_EN} track, GBDT has a distinct advantage in AUC.
        \item \textit{Developing Countries}: In developing countries, Multi-task learning generally leads in F1 score. However, their AUCs are again closely matched. This suggests that while the more complex model may predict certain classes better, both models rank their predictions similarly.
    \end{itemize}
\end{itemize}

\subsection{Fairness}

Understanding the model's performance in terms of fairness is especially crucial for platforms like Duolingo, where feedback directly influences learners' progress and motivation. Any disparity can lead to differing user experiences, potentially leading to decreased trust in the platform or even reduced learning outcomes for a particular user group.

\subsubsection{Fairness on Client}
\begin{figure}[ht]
    \centering
    \begin{minipage}{0.45\textwidth}
        \centering
        \includegraphics[width=1\linewidth]{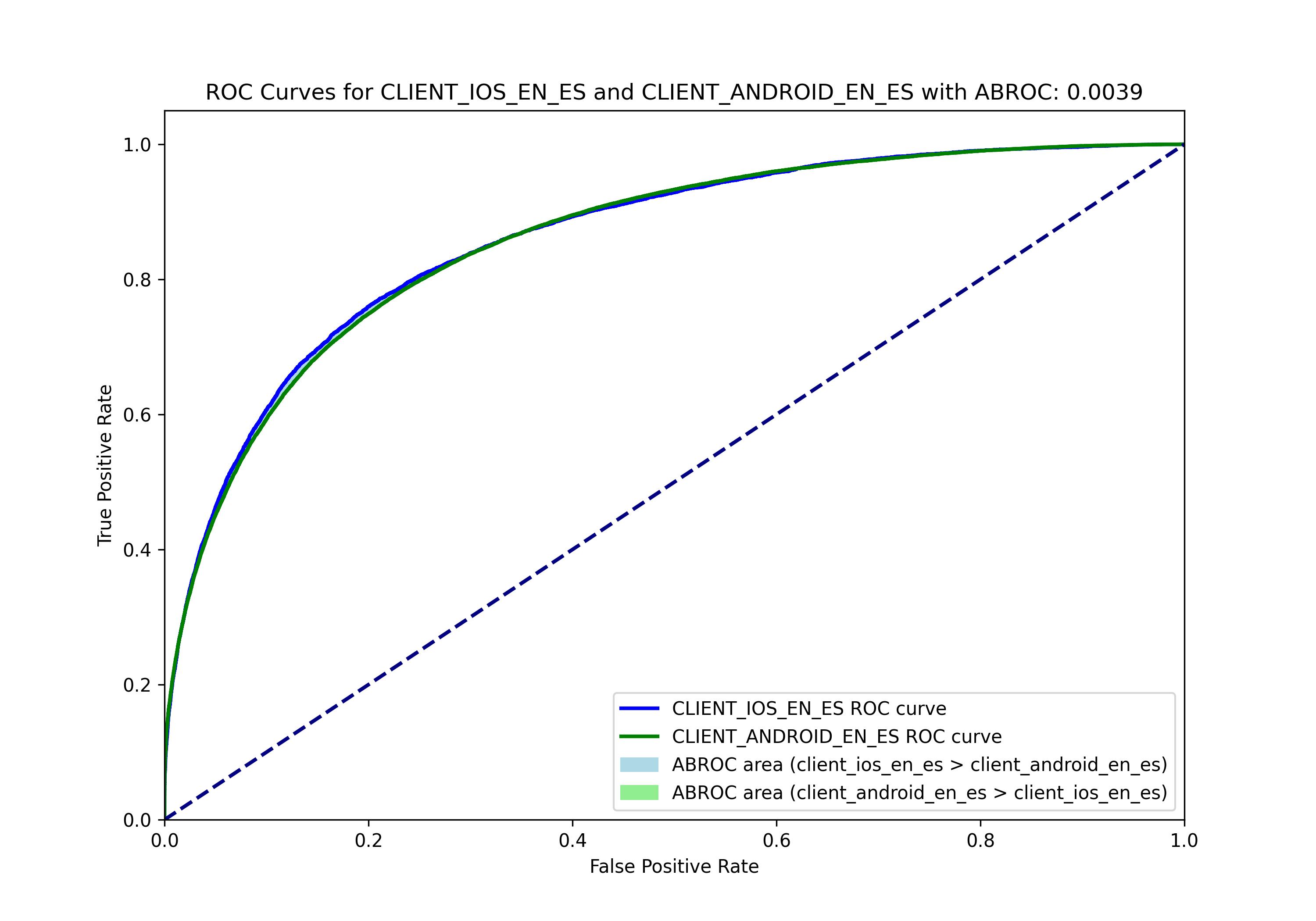}
        %Adjust the file type if needed
        \caption{Fairness between iOS and Android in \texttt{en\_es} track for GBDT}
        \label{fig:client GBDT}
    \end{minipage}\hfill
    \begin{minipage}{0.45\textwidth}
        \centering
        \includegraphics[width=1\linewidth]{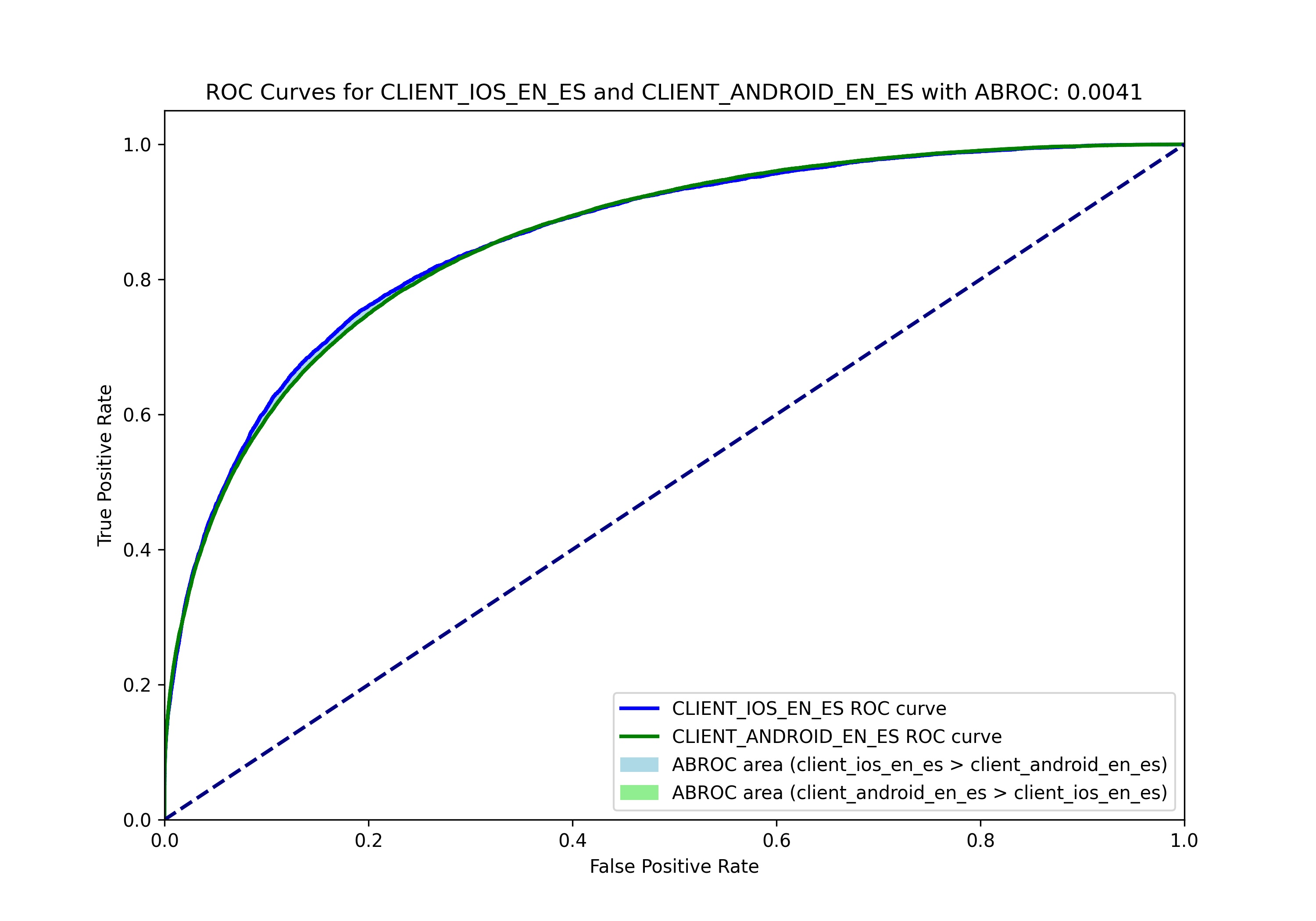} % Adjust the file type if needed
        \caption{Fairness between iOS and Android in \texttt{en\_es} track for Multi-task learning}
        \label{fig:client Multi-task}
    \end{minipage}
\end{figure}

Figures \ref{fig:client GBDT} shows the ABROCA between iOS and Android in \texttt{en\_es} track for GBDT. Figure \ref{fig:client Multi-task} shows the ABROCA between iOS and Android in \texttt{en\_es} track for Multi-task learning.

To better elucidate the differences in ABROCA with respect to client-side performance between GBDT and Multi-task learning, we present the following Table \ref{tab:fairnessClient}.

\begin{table}[h]
    \centering
    \caption{Fairness on Client}
    \label{tab:fairnessClient}
    \begin{tabular}{lllcc}
        \toprule
        Client 1 & Client 2 & Track & GBDT(ABROCA) & Multi-task learning(ABROCA) \\
        \midrule
        Android & Web & en\_es & 0.0226 & 0.0202 \\
        ios & Android & en\_es & 0.0039 & 0.0041 \\
        ios & Web & en\_es & 0.0236 & 0.0198 \\
        Android & Web & es\_en & 0.0307 & 0.0406 \\
        ios & Android & es\_en & 0.0082 & 0.0078 \\
        ios & Web & es\_en & 0.0227 & 0.0318 \\
        Android & Web & fr\_en & 0.0333 & 0.0073 \\
        ios & Android & fr\_en & 0.002 & 0.0091 \\
        ios & Web & fr\_en & 0.0341 & 0.0138 \\
        \bottomrule
    \end{tabular}
\end{table}

Overall, the ABROCA on the Android and iOS platforms is noticeably smaller than in the other categories across the three linguistic tracks, highlighting the inherent similarity between the Android and iOS mobile platforms in the context of algorithmic fairness. Pertinently, both the GBDT and Multi-task learning algorithms exhibit discernible biases against Web users. Despite the notable that the ABROCA of the GBDT end is larger than that of the Multi-task learning, indicating a somewhat larger discriminatory impact against non-mobile users, it's imperative to note that neither algorithm provides an entirely impartial solution when considering platform-based disparities. Thus, while differences in the extent of bias exist between the two algorithms, our analysis leads us to the indispensable conclusion that both GBDT and Multi-task learning algorithms present prejudicial outcomes against non-mobile (Web) users, thereby necessitating a further exploration into mechanisms that can mitigate such biases to ensure fair and equitable knowledge tracing across all platforms.

\subsubsection{Fairness on Country}
\begin{figure}[ht]
    \centering
    \begin{minipage}{0.45\textwidth}
        \centering
        \includegraphics[width=1\linewidth]{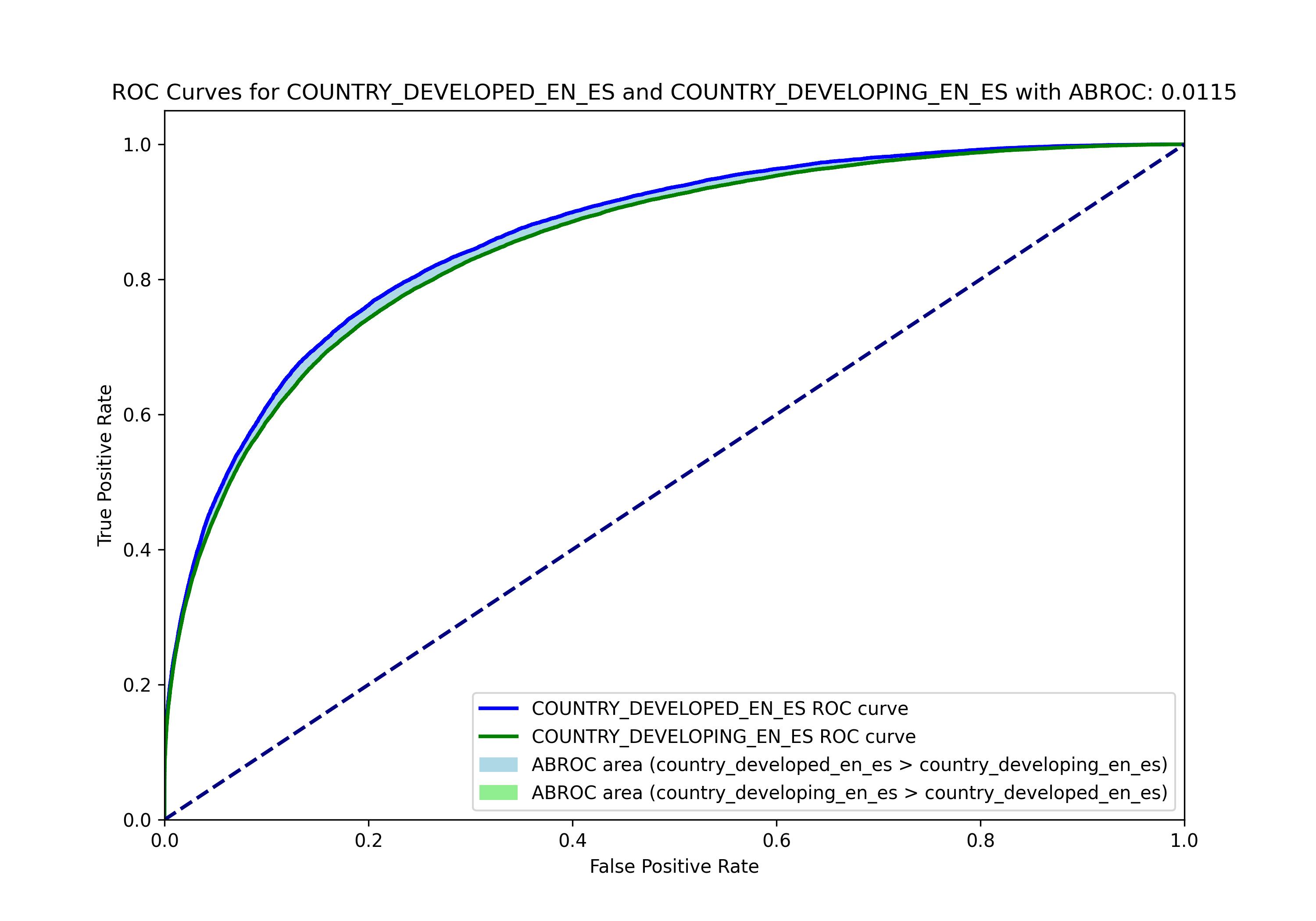} % Adjust the file type if needed
        \caption{Fairness between developed country and developing country in \texttt{en\_es} track for GBDT}
        \label{fig:country GBDT}
    \end{minipage}\hfill
    \begin{minipage}{0.45\textwidth}
        \centering
        \includegraphics[width=1\linewidth]{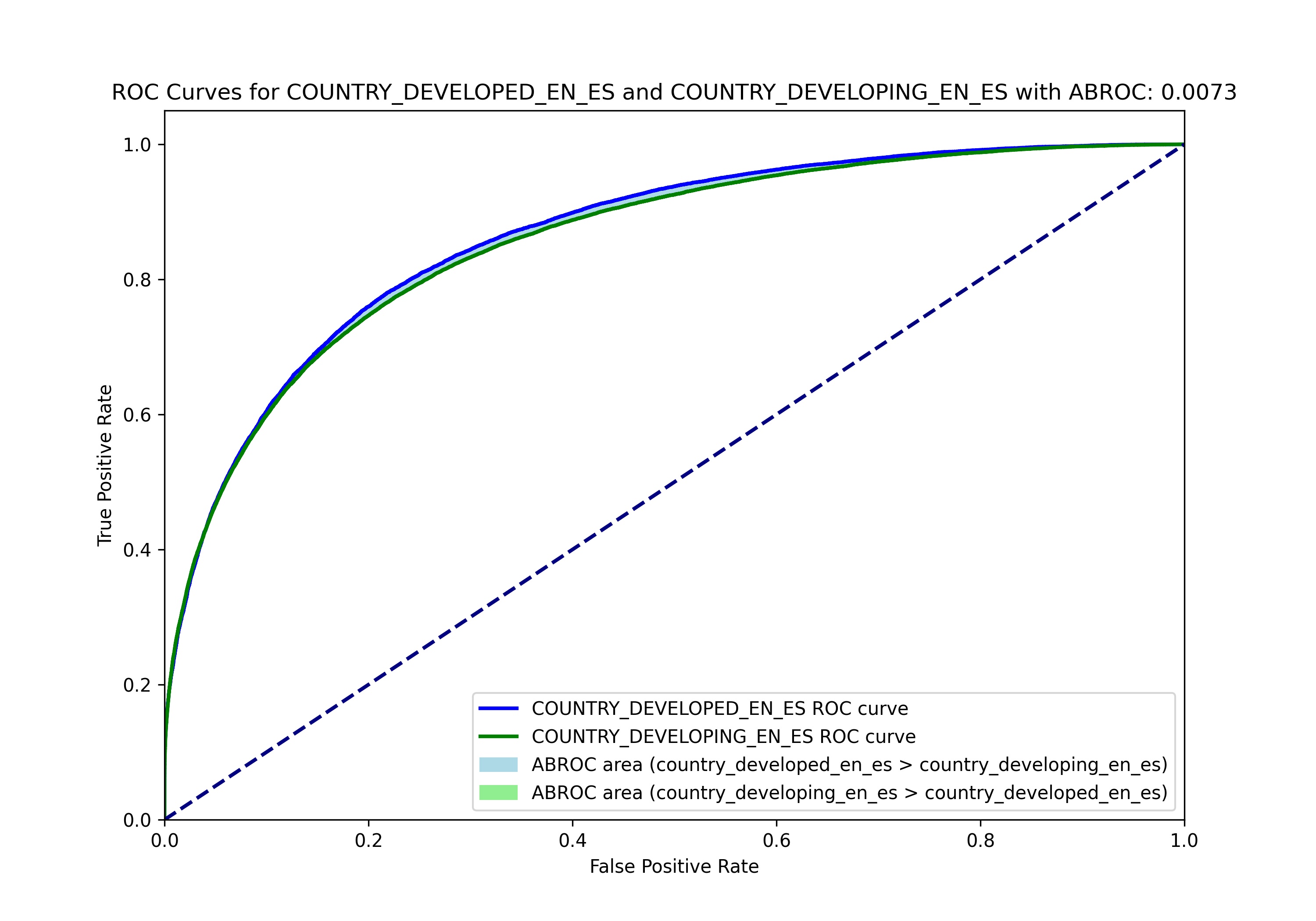} % Adjust the file type if needed
        \caption{Fairness between developed country and developing country in \texttt{en\_es} track for Multi-task learning}
        \label{fig:country Multi-task}
    \end{minipage}
\end{figure}

Figures \ref{fig:country GBDT} shows the ABROCA between developed country and developing country in \texttt{en\_es} track for GBDT. Figure \ref{fig:country Multi-task} shows the ABROCA between developed country and developing country in \texttt{en\_es} track for Multi-task learning.
To better elucidate the differences in ABROCA with respect to country-side performance between GBDT and Multi-task learning, we present the following table \ref{tab:fairnessCountryDeveloped}.

\begin{table}[h]
    \centering
    \caption{Fairness on Country(developed country vs developing country)}
    \label{tab:fairnessCountryDeveloped}
    \begin{tabular}{lcc}
        \toprule
        Track & GBDT(ABROCA) & Multi-task learning(ABROCA) \\
        \midrule
        en\_es & 0.0115 & 0.0073 \\
        es\_en & 0.0154 & 0.0139 \\
        fr\_en & 0.0026 & 0.0062 \\
        \bottomrule
    \end{tabular}
\end{table}
We can observe that Multi-task learning demonstrates greater fairness than GBDT in the  \texttt{en\_es} and  
 \texttt{es\_en} tracks, while it exhibits less fairness in the  \texttt{fr\_en} track. This indicates that for the  \texttt{fr\_en} track, GBDT possesses superior accuracy and fairness. Conversely, for the  \texttt{en\_es} and  \texttt{es\_en} tracks, Multi-task learning holds a greater advantage. Viewing holistically, GBDT exhibits more bias against developing countries compared to Multi-task learning.

 \section{Discussion and Conclusion}
 This paper examines the fairness and accuracy of two knowledge tracing algorithms, GBDT and Multi-task learning, in the domain of second language acquisition across three linguistic tracks: \texttt{en\_es}, \texttt{es\_en}, and \texttt{fr\_en}. Our comprehensive analysis reveals key insights into the interplay between algorithm choice and its impact on fairness and accuracy.

Firstly, our findings underscore that deep learning, exemplified here by Multi-task learning, tends to be more apt for knowledge tracing based on second language acquisition than machine learning algorithms like GBDT, owing predominantly to its higher accuracy and fairness, especially in the \texttt{en\_es} and \texttt{es\_en} tracks. Secondly, a discernible discriminatory tendency against non-mobile end-users was observed in both machine learning and deep learning algorithms. Furthermore, our third finding points towards a more pronounced bias inherent in machine learning algorithms, particularly towards developing countries, as opposed to deep learning algorithms.

To assess the potential algorithmic bias towards developing countries and users of web and android platforms (excluding iOS), we employed ABROCA as our fairness measurement tool. The fairness and accuracy analysis starkly illuminated the superiority of Multi-task learning in terms of algorithmic fairness compared to the GBDT algorithm.

Despite Multi-task learning overwhelmingly emerging as the preferable option due to its enhanced accuracy and algorithmic fairness, our fourth finding elucidates a nuanced approach to algorithm selection across different linguistic tracks. While Multi-task learning would be the optimal choice for \texttt{en\_es} and \texttt{es\_en} due to its fairness and predictive proficiency, for \texttt{fr\_en} track, the GBDT algorithm is recommended due to its superior accuracy and fairness, despite Multi-task learning’s broader applicability and success.

In summary, while Multi-task learning broadly demonstrates enviable aptitude in knowledge tracing, especially when balancing fairness and accuracy, the GBDT algorithm's selective applicability, such as in \texttt{fr\_en} track scenarios, indicates that a tailored, context-specific approach to algorithm selection offers an optimal strategy in ensuring equitable and accurate knowledge tracing across varied linguistic and platform-specific contexts.

\section{Limitations}
We acknowledge several limitations in our study. Firstly, our analysis was confined to a single predictive task: knowledge tracing in second language acquisition, utilizing only the Duolingo dataset. This limitation restricts the generalizability of our findings across different datasets and contexts. To enhance the universality and robustness of our results, future studies should explore a broader range of machine learning and neural network algorithms and apply them to other predictive tasks in varied datasets. Additionally, our examination of demographic differences was limited to two specific student groups: Client and Country. Future research should extend this analysis to include other demographic factors, such as age and gender, to provide a more comprehensive evaluation of algorithmic fairness across diverse student populations.

%%%%%%%%%%%%%%%%%%%%%%%%%%%%%%%%%%%%%%%%%%%%%%%%%%%%%%%%%%%%

\appendix

\section{Appendix / supplemental material}

\textbf{All the graphs related to fairness on Client are listed below}

\begin{figure}[htbp]
\centering
\begin{minipage}[t]{0.45\textwidth}
    \centering
    \includegraphics[width=\linewidth]{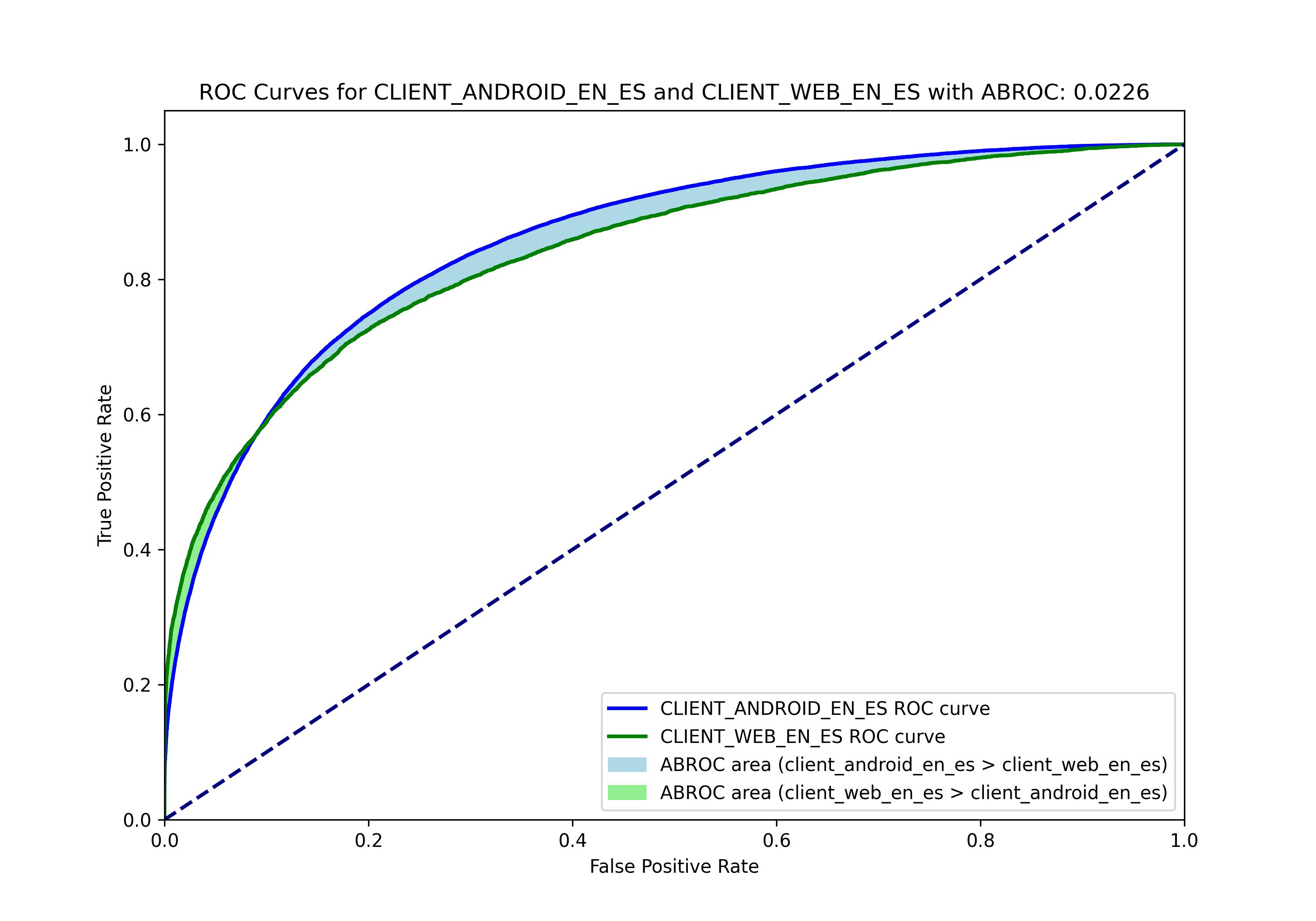}
    \caption{Fairness between android and web in \texttt{en\_es} track for GBDT}
    \label{fig:client_android_en_es_VS_client_web_en_es_ROC_ABROC}
\end{minipage}\hfill
\begin{minipage}[t]{0.45\textwidth}
    \centering
    \includegraphics[width=\linewidth]{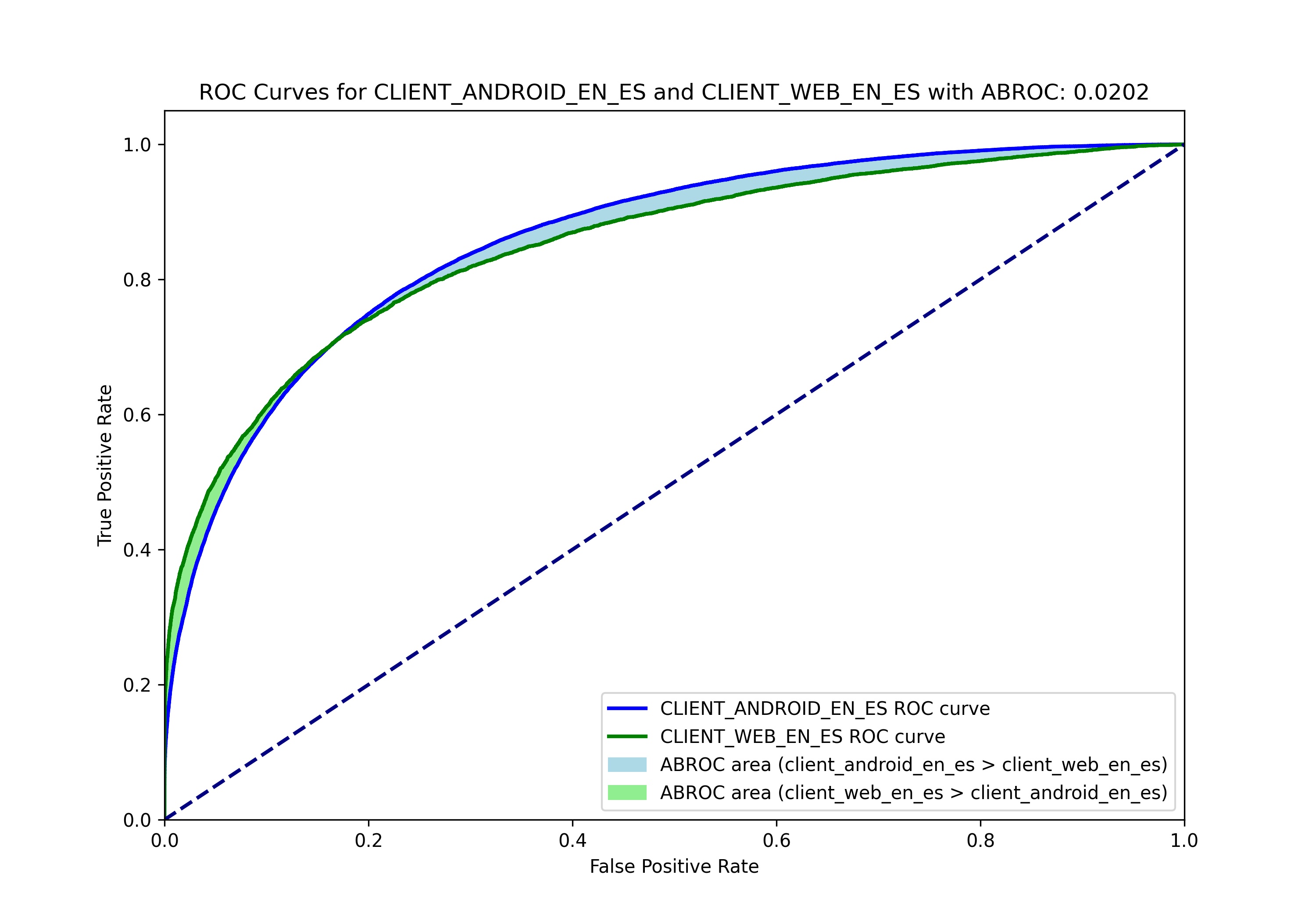}
    \caption{Fairness between android and web in \texttt{en\_es} track for Multi-task learning}
    \label{fig:client_android_en_es_VS_client_web_en_es_ROC_ABROC}
\end{minipage}\hfill
\end{figure}

\begin{figure}[htbp]
\centering
\begin{minipage}[t]{0.45\textwidth}
    \centering
    \includegraphics[width=\linewidth]{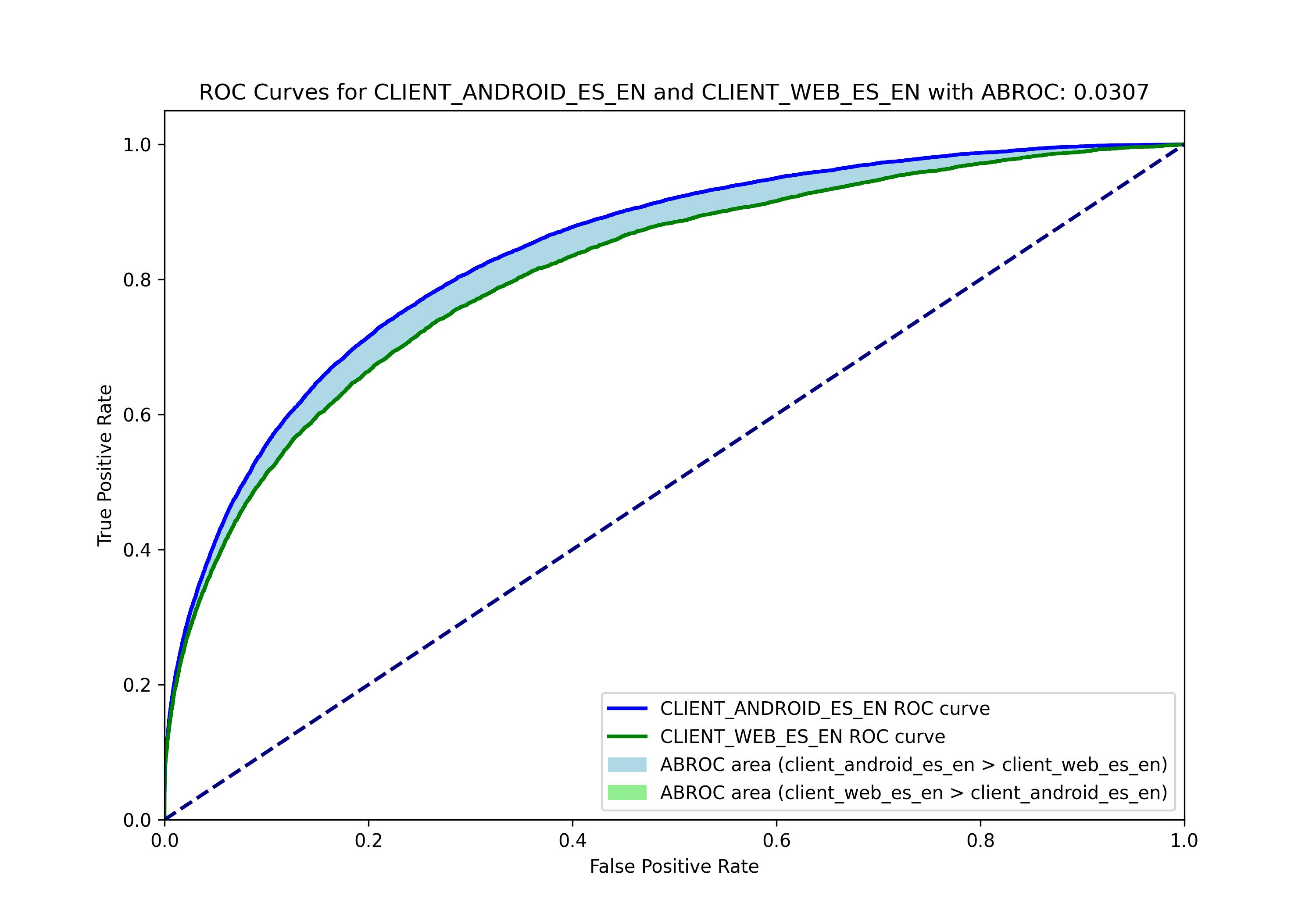}
    \caption{Fairness between android and web in \texttt{es\_en} track for GBDT}
    \label{fig:client_android_es_en_VS_client_web_es_en_ROC_ABROC}
\end{minipage}\hfill
\begin{minipage}[t]{0.45\textwidth}
    \centering
    \includegraphics[width=\linewidth]{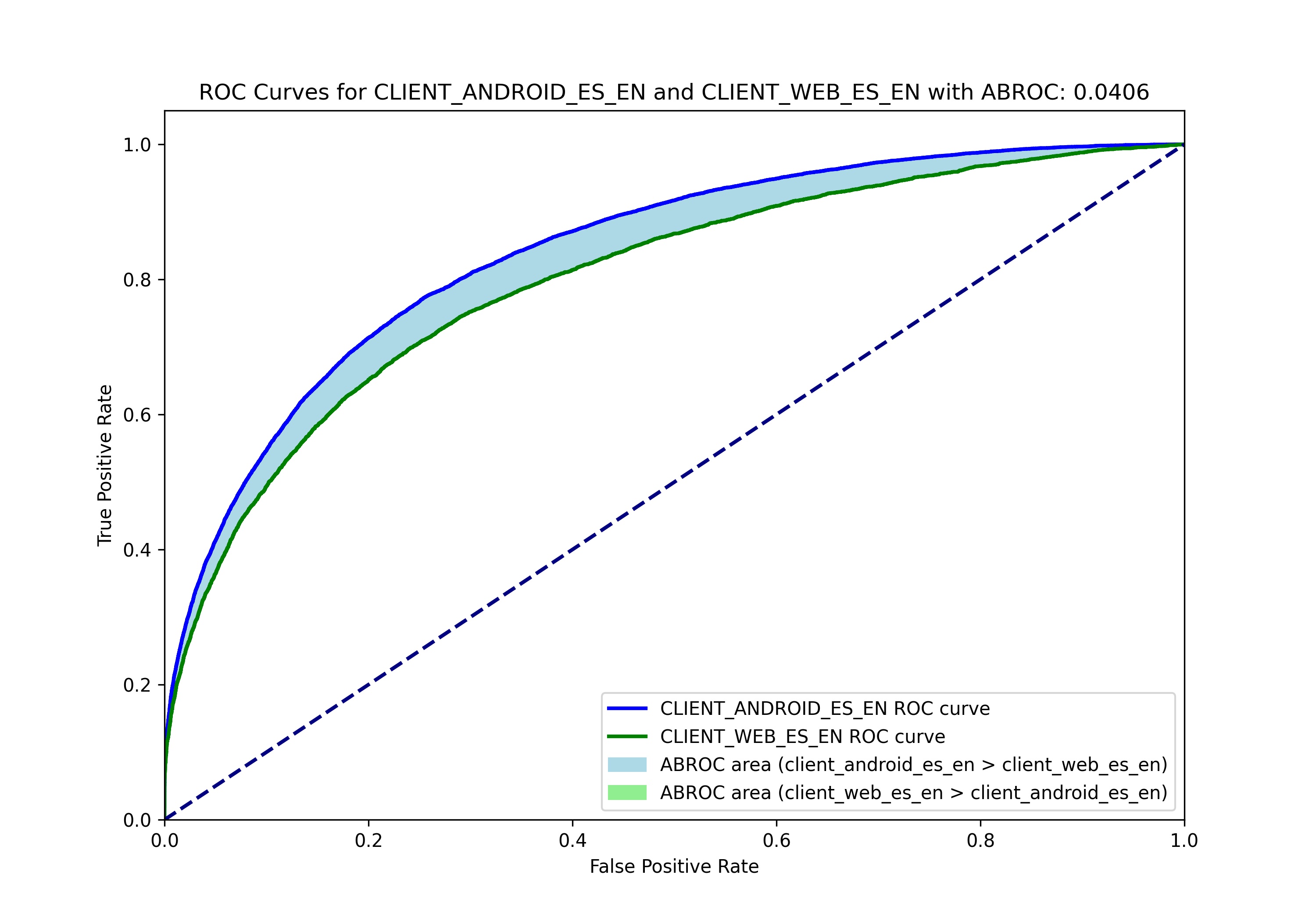}
    \caption{Fairness between android and web in \texttt{es\_en} track for Multi-task learning}
    \label{fig:client_android_es_en_VS_client_web_es_en_ROC_ABROC}
\end{minipage}\hfill
\end{figure}

\begin{figure}[htbp]
\centering
\begin{minipage}[t]{0.45\textwidth}
    \centering
    \includegraphics[width=\linewidth]{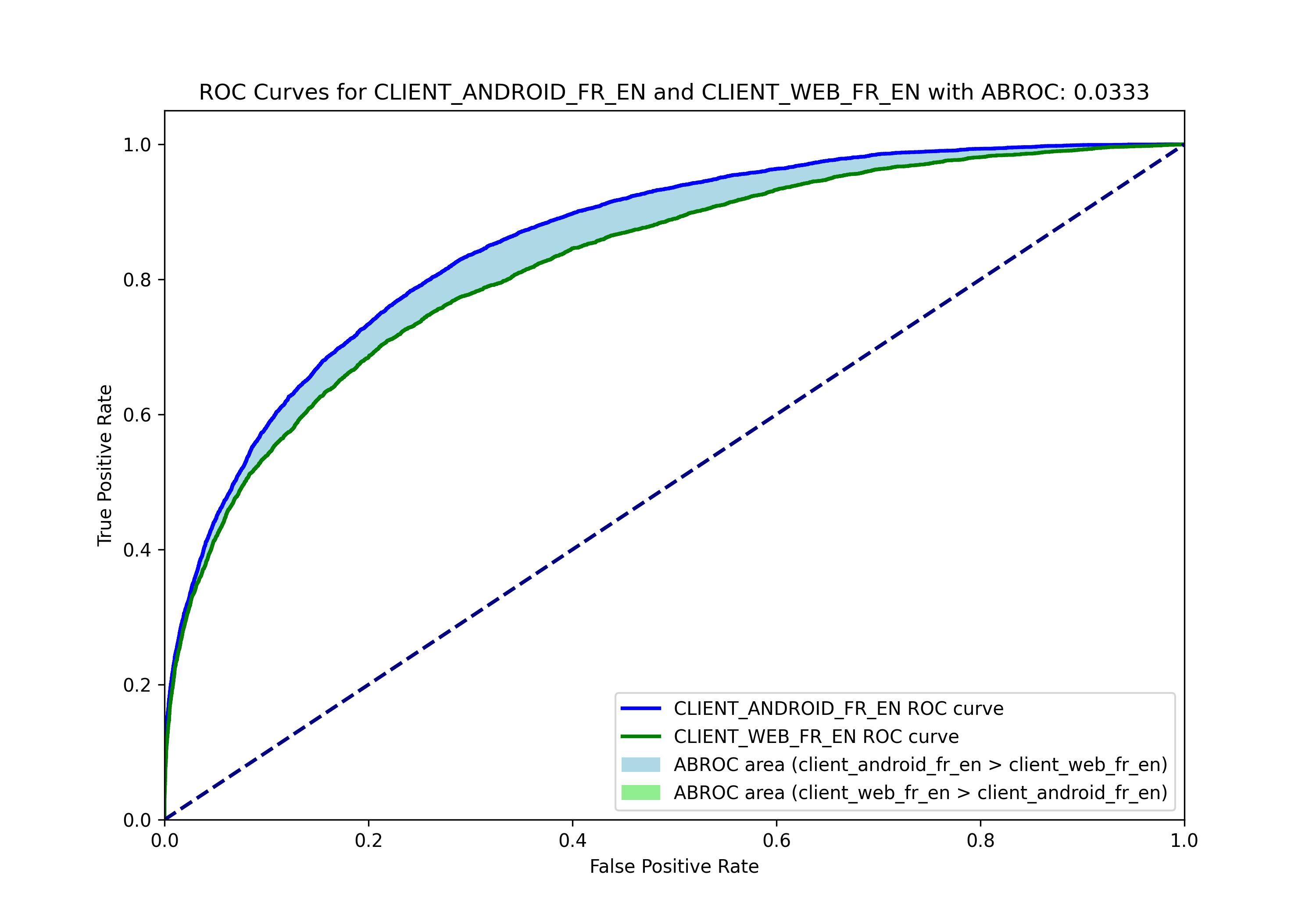}
    \caption{Fairness between android and web in \texttt{fr\_en} track for GBDT}
    \label{fig:client_android_fr_en_VS_client_web_fr_en_ROC_ABROC}
\end{minipage}\hfill
\begin{minipage}[t]{0.45\textwidth}
    \centering
    \includegraphics[width=\linewidth]{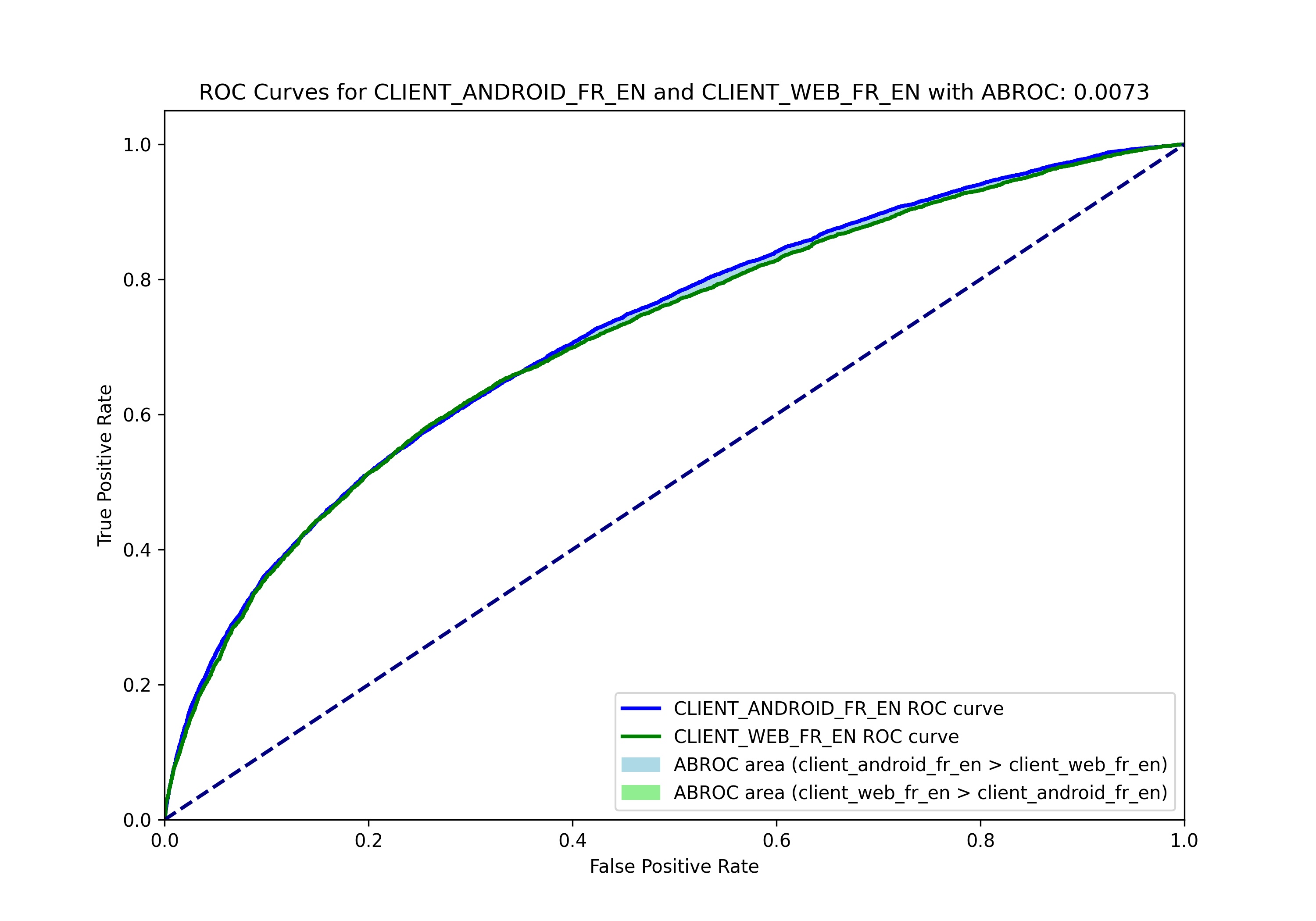}
    \caption{Fairness between android and web in \texttt{fr\_en} track for Multi-task learning}
    \label{fig:client_android_fr_en_VS_client_web_fr_en_ROC_ABROC}
\end{minipage}\hfill
\end{figure}

\begin{figure}[htbp]
\centering
\begin{minipage}[t]{0.45\textwidth}
    \centering
    \includegraphics[width=\linewidth]{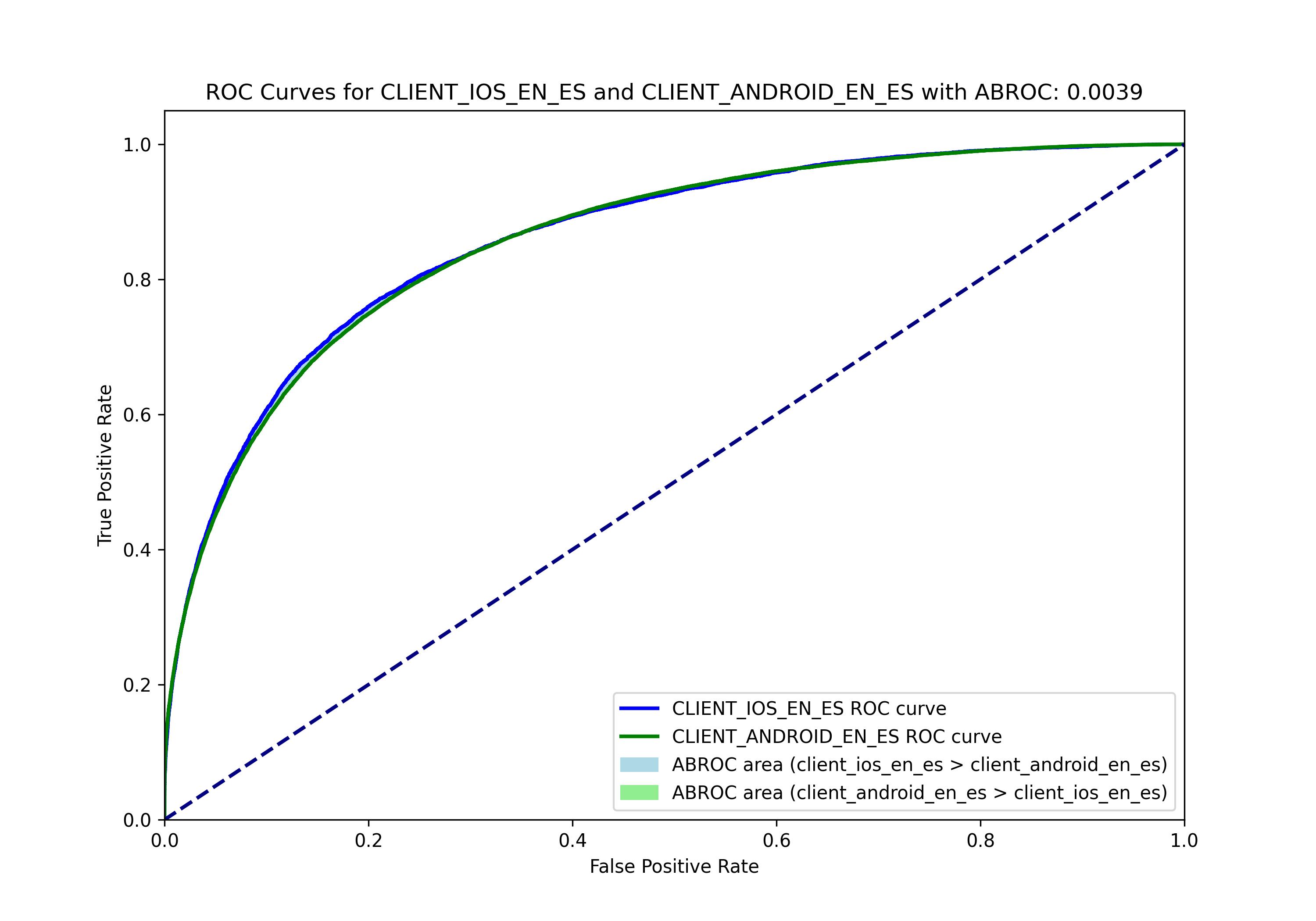}
    \caption{Fairness between ios and android in \texttt{en\_es} track for GBDT}
    \label{fig:client_ios_en_es_VS_client_android_en_es_ROC_ABROC}
\end{minipage}\hfill
\begin{minipage}[t]{0.45\textwidth}
    \centering
    \includegraphics[width=\linewidth]{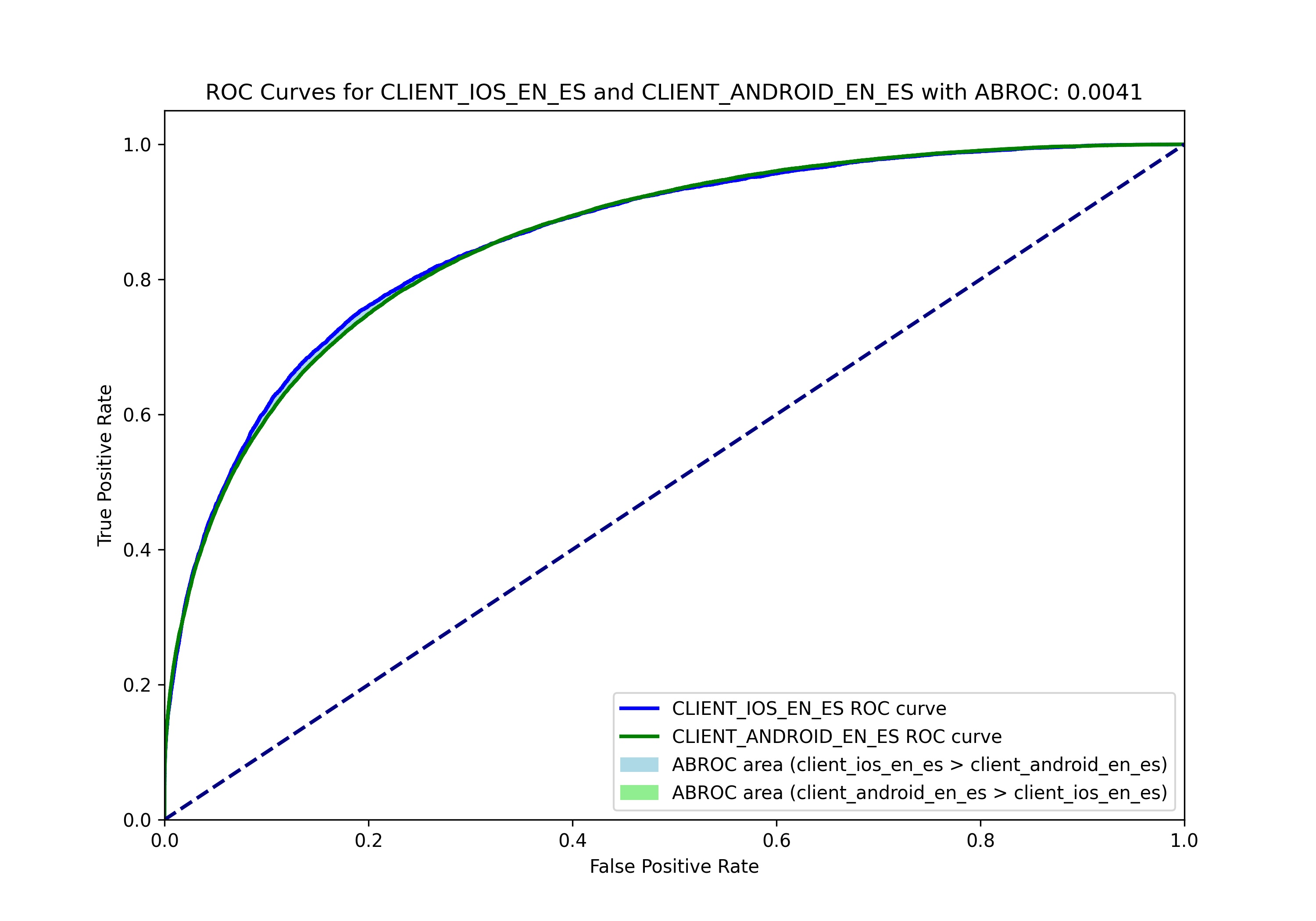}
    \caption{Fairness between ios and android in \texttt{en\_es} track for Multi-task learning}
    \label{fig:client_ios_en_es_VS_client_android_en_es_ROC_ABROC}
\end{minipage}\hfill
\end{figure}

\begin{figure}[htbp]
\centering
\begin{minipage}[t]{0.45\textwidth}
    \centering
    \includegraphics[width=\linewidth]{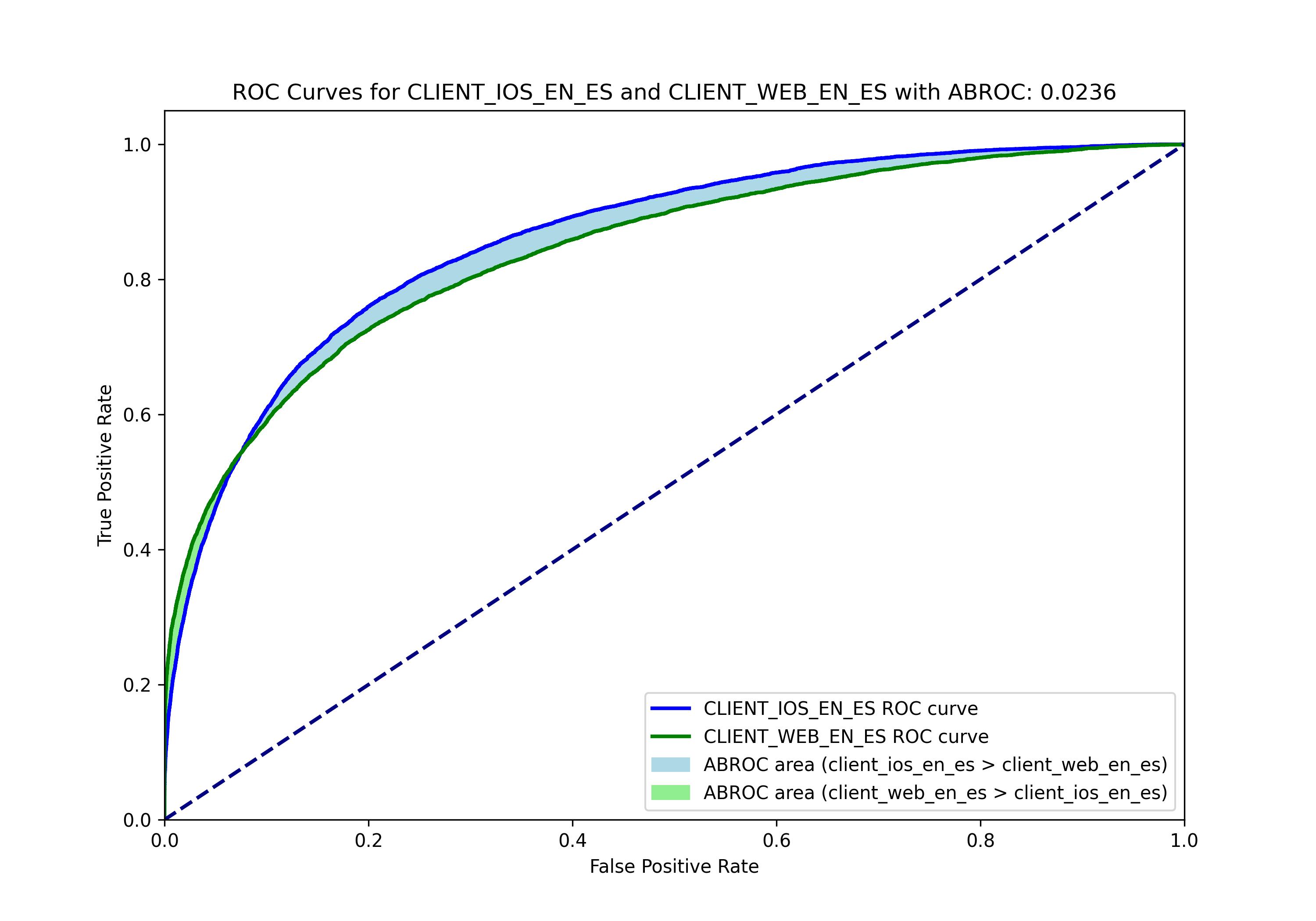}
    \caption{Fairness between ios and web in \texttt{en\_es} track for GBDT}
    \label{fig:client_ios_en_es_VS_client_web_en_es_ROC_ABROC}
\end{minipage}\hfill
\begin{minipage}[t]{0.45\textwidth}
    \centering
    \includegraphics[width=\linewidth]{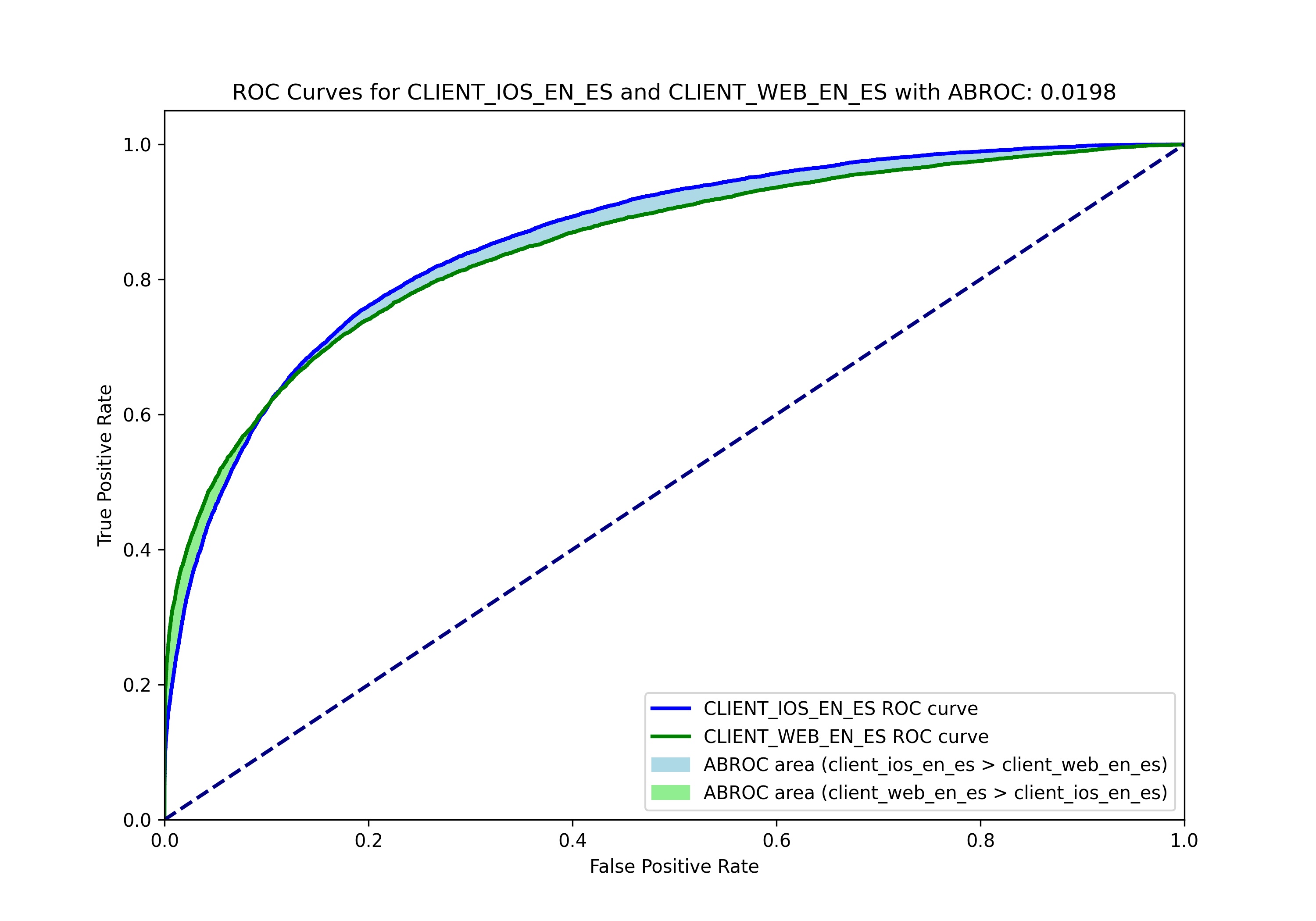}
    \caption{Fairness between ios and web in \texttt{en\_es} track for Multi-task learning}
    \label{fig:client_ios_en_es_VS_client_web_en_es_ROC_ABROC}
\end{minipage}\hfill
\end{figure}

\begin{figure}[htbp]
\centering
\begin{minipage}[t]{0.45\textwidth}
    \centering
    \includegraphics[width=\linewidth]{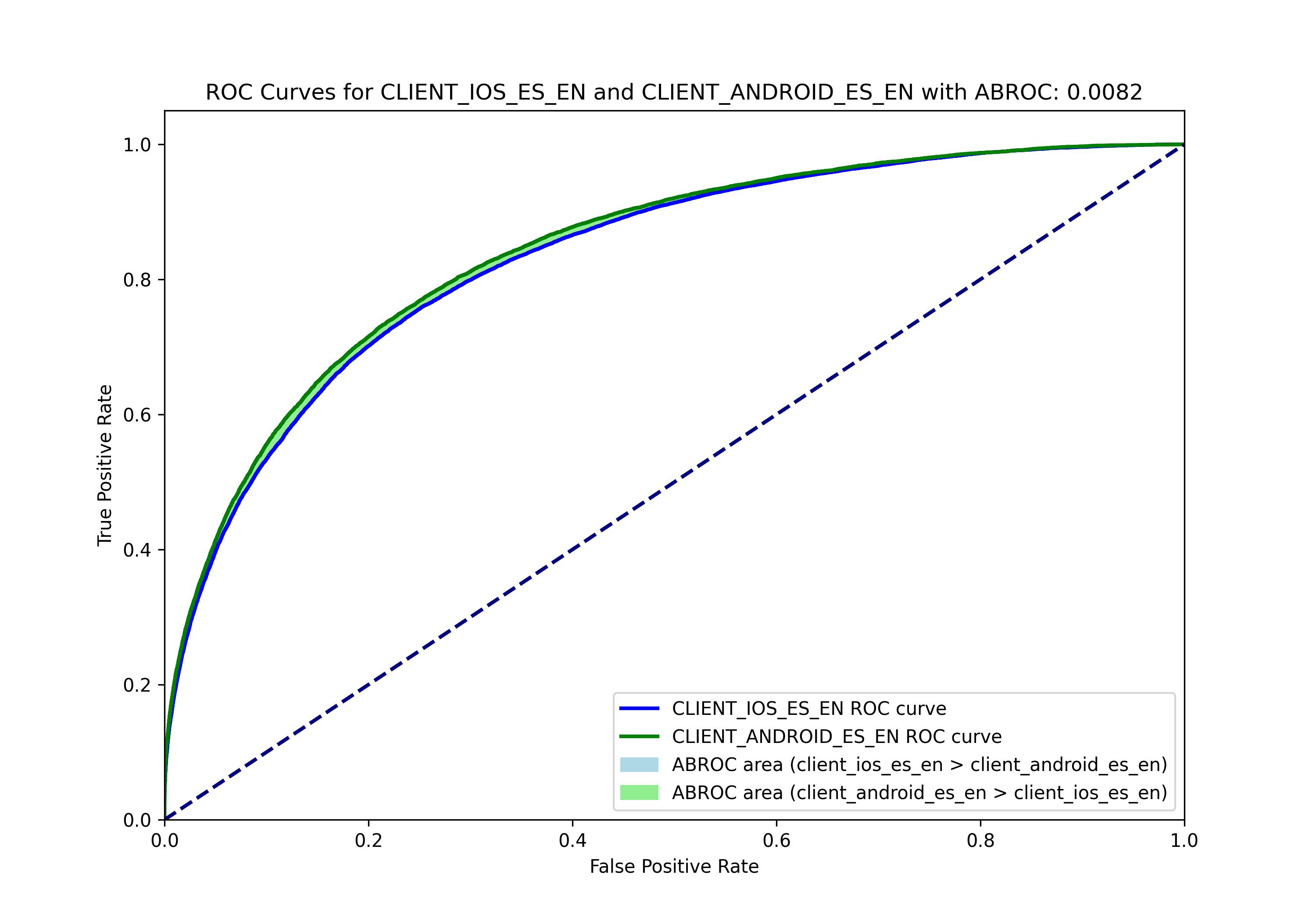}
    \caption{Fairness between ios and android in \texttt{es\_en} track for GBDT}
    \label{fig:client_ios_es_en_VS_client_android_es_en_ROC_ABROC}
\end{minipage}\hfill
\begin{minipage}[t]{0.45\textwidth}
    \centering
    \includegraphics[width=\linewidth]{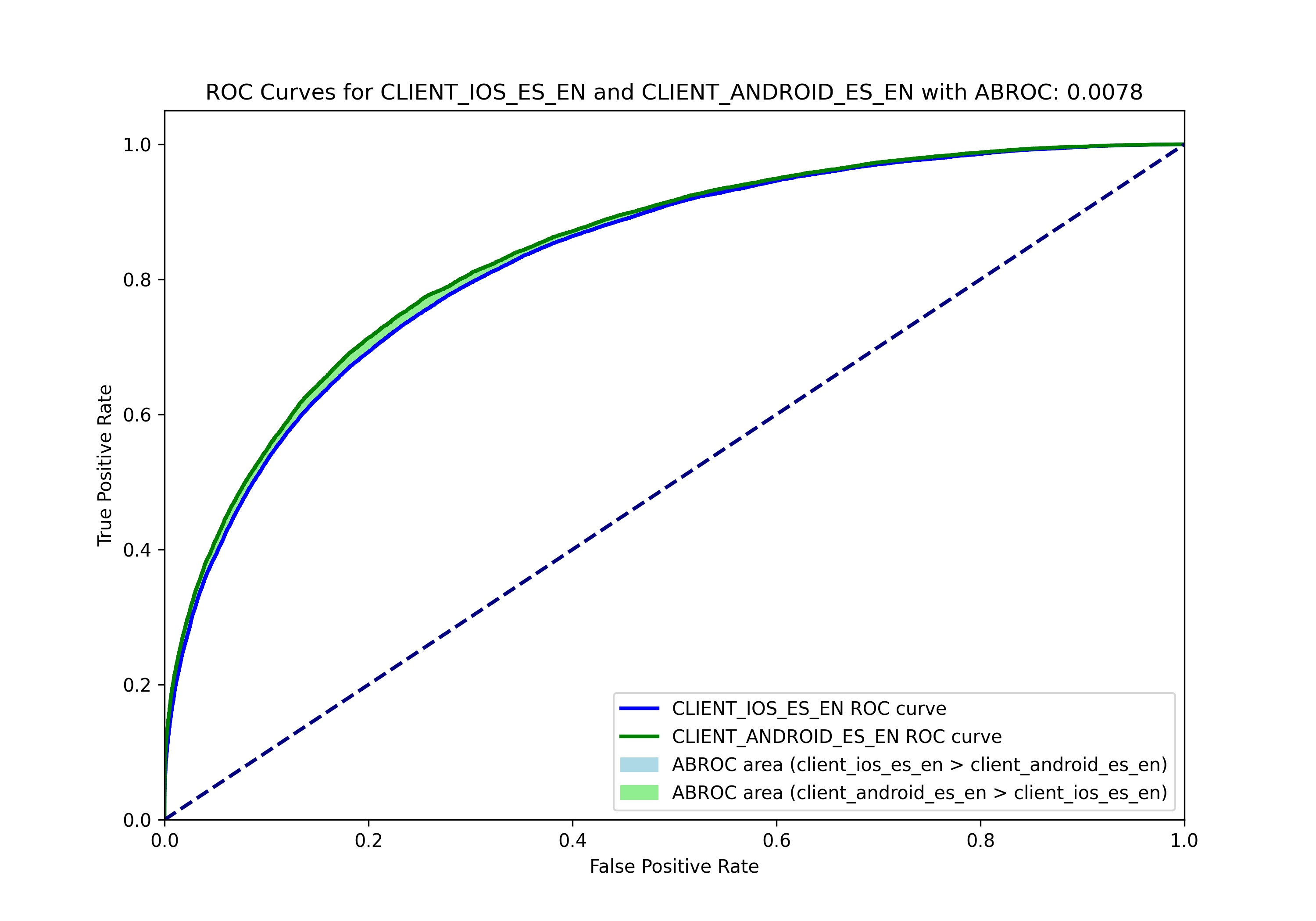}
    \caption{Fairness between ios and android in \texttt{es\_en} track for Multi-task learning}
    \label{fig:client_ios_es_en_VS_client_android_es_en_ROC_ABROC}
\end{minipage}\hfill
\end{figure}

\begin{figure}[htbp]
\centering
\begin{minipage}[t]{0.45\textwidth}
    \centering
    \includegraphics[width=\linewidth]{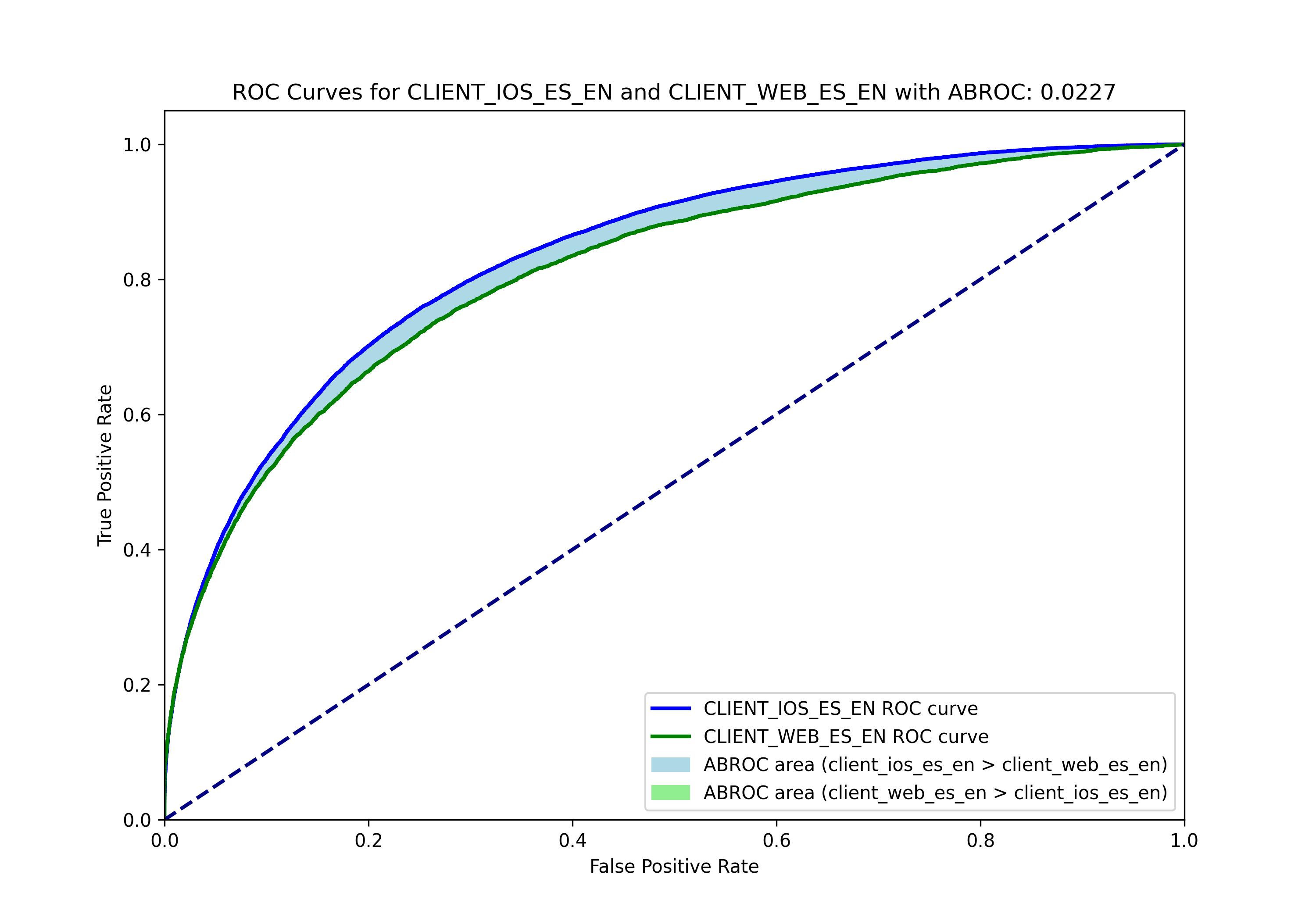}
    \caption{Fairness between ios and web in \texttt{es\_en} track for GBDT}
    \label{fig:client_ios_es_en_VS_client_web_es_en_ROC_ABROC}
\end{minipage}\hfill
\begin{minipage}[t]{0.45\textwidth}
    \centering
    \includegraphics[width=\linewidth]{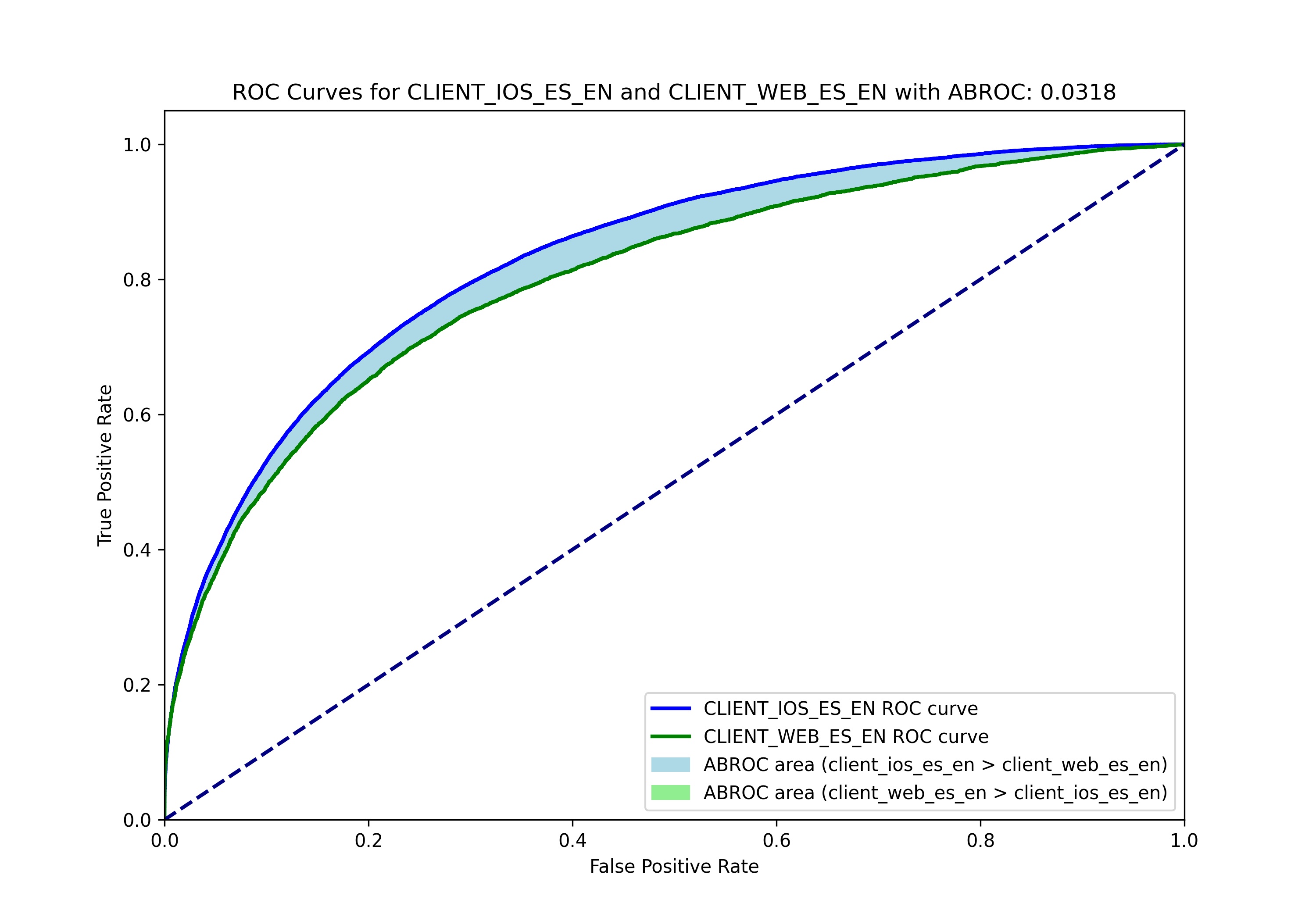}
    \caption{Fairness between ios and web in \texttt{es\_en} track for Multi-task learning}
    \label{fig:client_ios_es_en_VS_client_web_es_en_ROC_ABROC}
\end{minipage}\hfill
\end{figure}

\newpage
\begin{figure}[htbp]
\centering
\begin{minipage}[t]{0.45\textwidth}
    \centering
    \includegraphics[width=\linewidth]{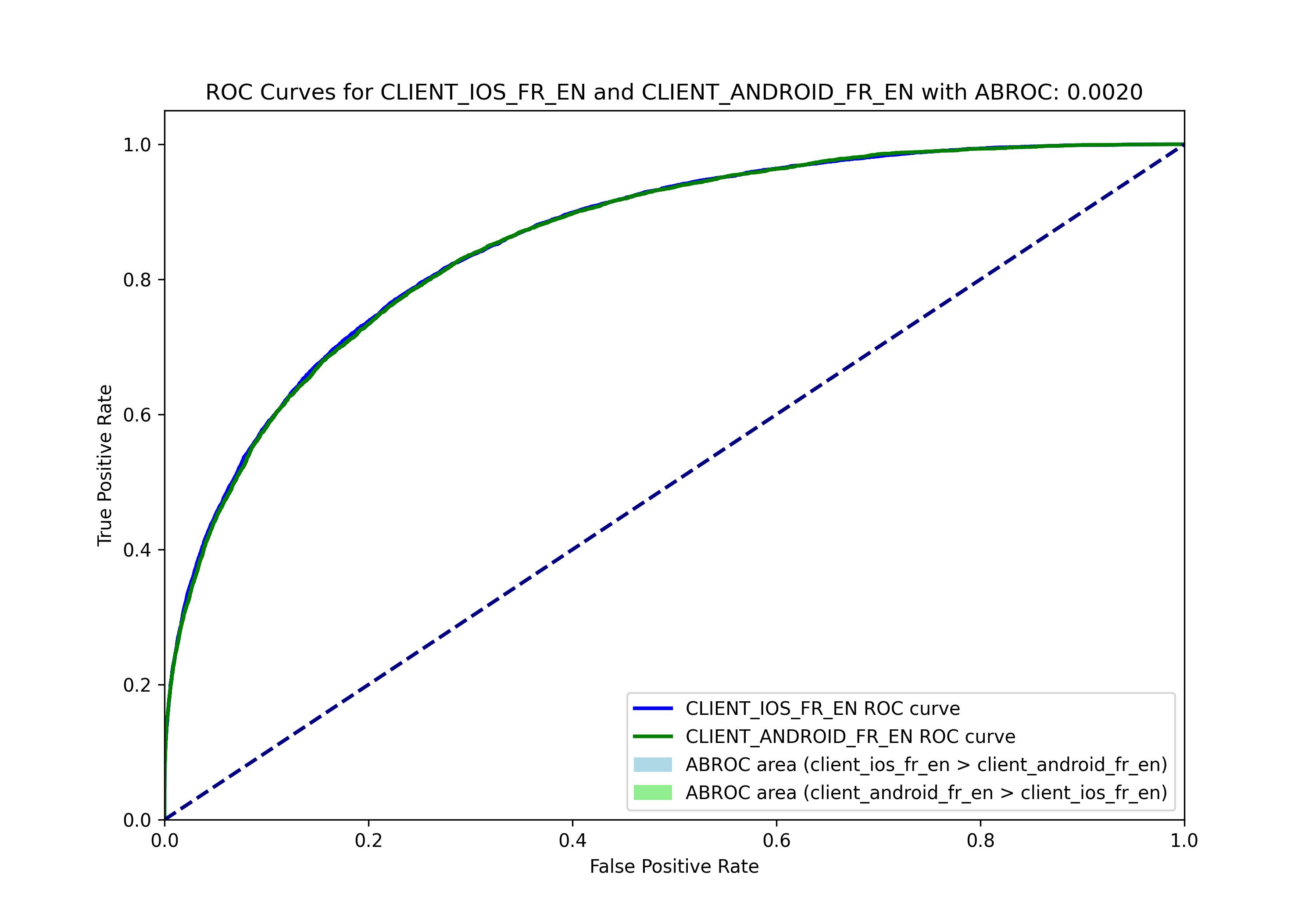}
    \caption{Fairness between ios and android in \texttt{fr\_en} track for GBDT}
    \label{fig:client_ios_fr_en_VS_client_android_fr_en_ROC_ABROC}
\end{minipage}\hfill
\begin{minipage}[t]{0.45\textwidth}
    \centering
    \includegraphics[width=\linewidth]{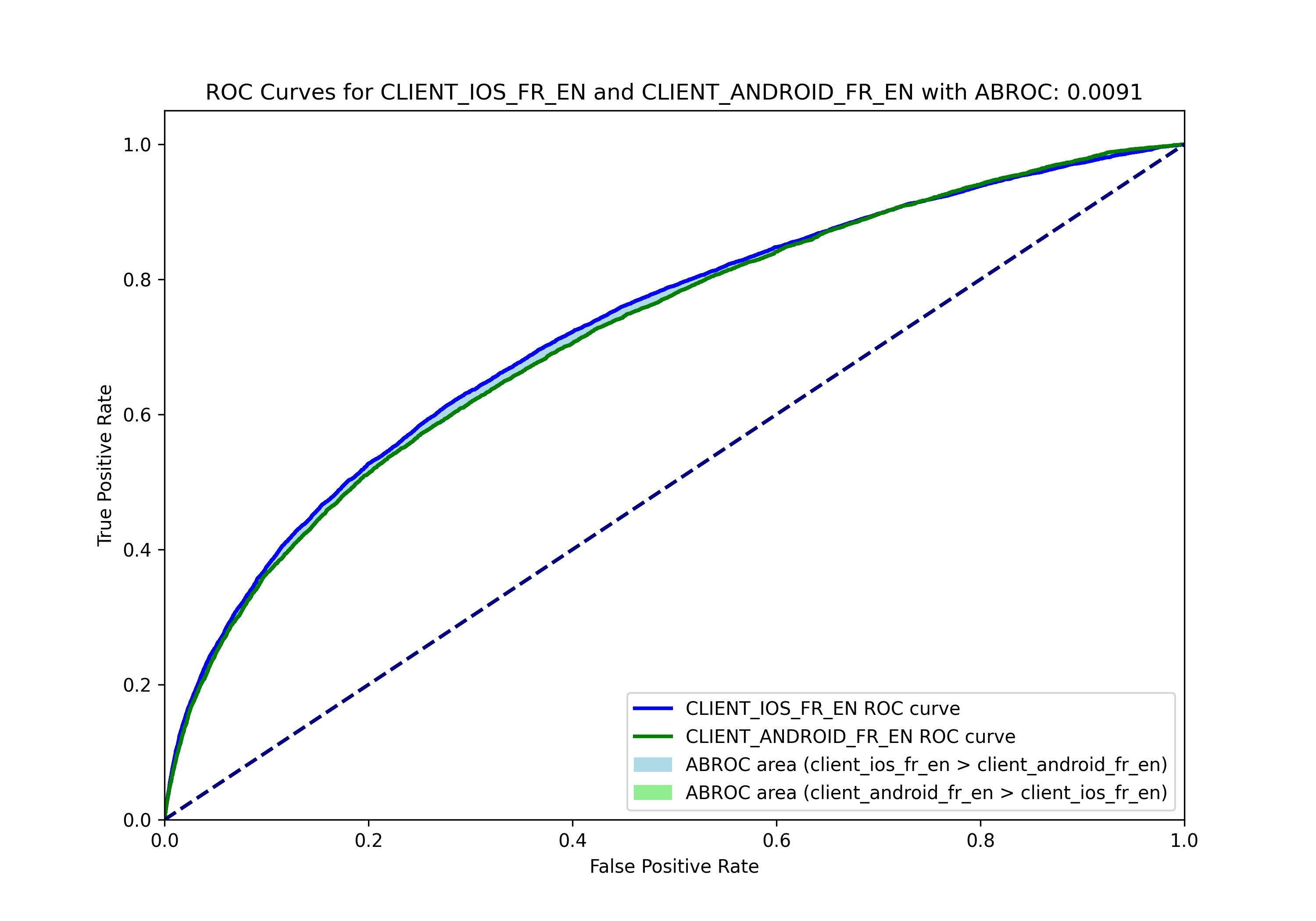}
    \caption{Fairness between ios and android in \texttt{fr\_en} track for Multi-task learning}
    \label{fig:client_ios_fr_en_VS_client_android_fr_en_ROC_ABROC}
\end{minipage}\hfill
\end{figure}

\begin{figure}[htbp]
\centering
\begin{minipage}[t]{0.45\textwidth}
    \centering
    \includegraphics[width=\linewidth]{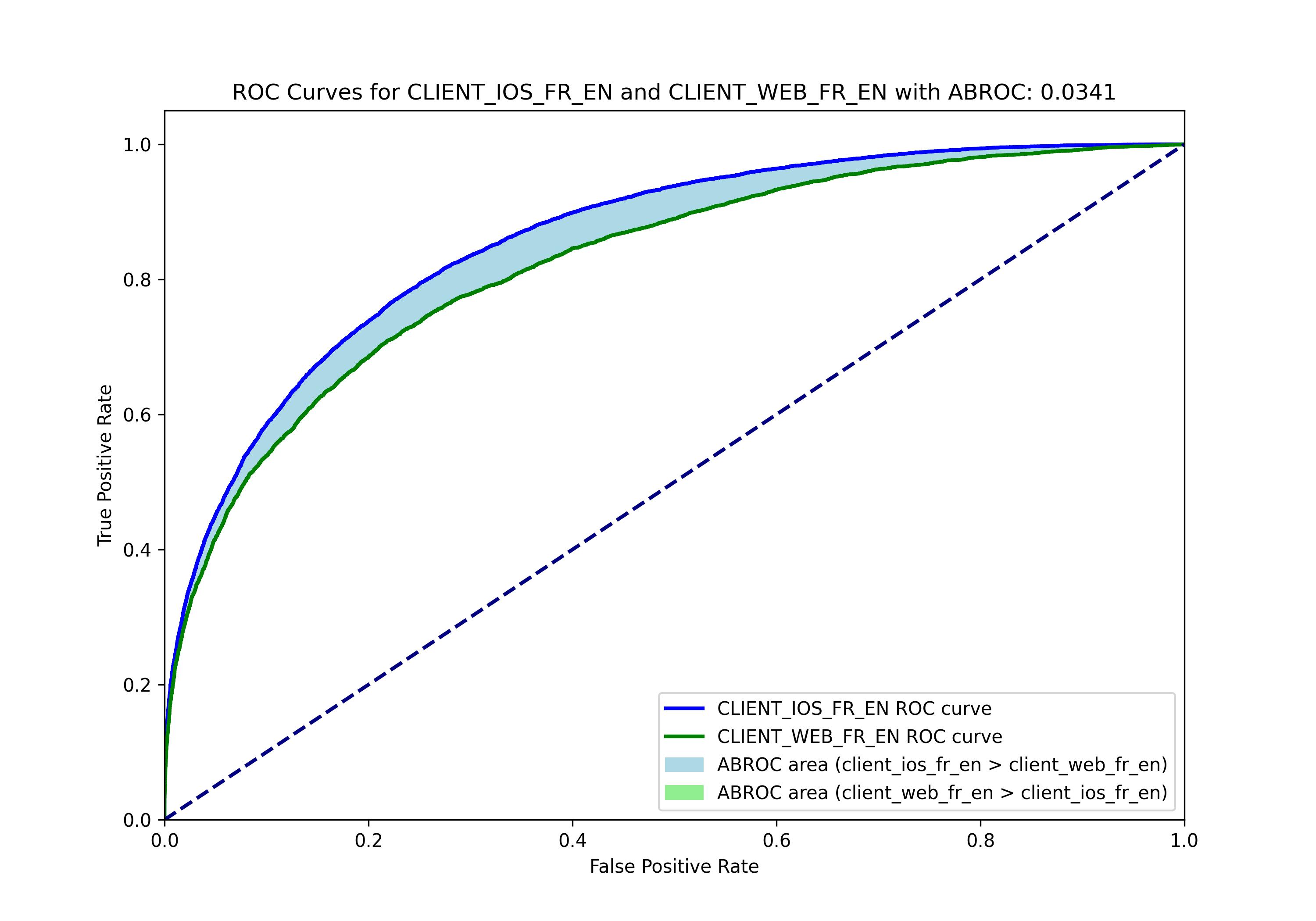}
    \caption{Fairness between ios and web in \texttt{fr\_en} track for GBDT}
    \label{fig:client_ios_fr_en_VS_client_web_fr_en_ROC_ABROC}
\end{minipage}\hfill
\begin{minipage}[t]{0.45\textwidth}
    \centering
    \includegraphics[width=\linewidth]{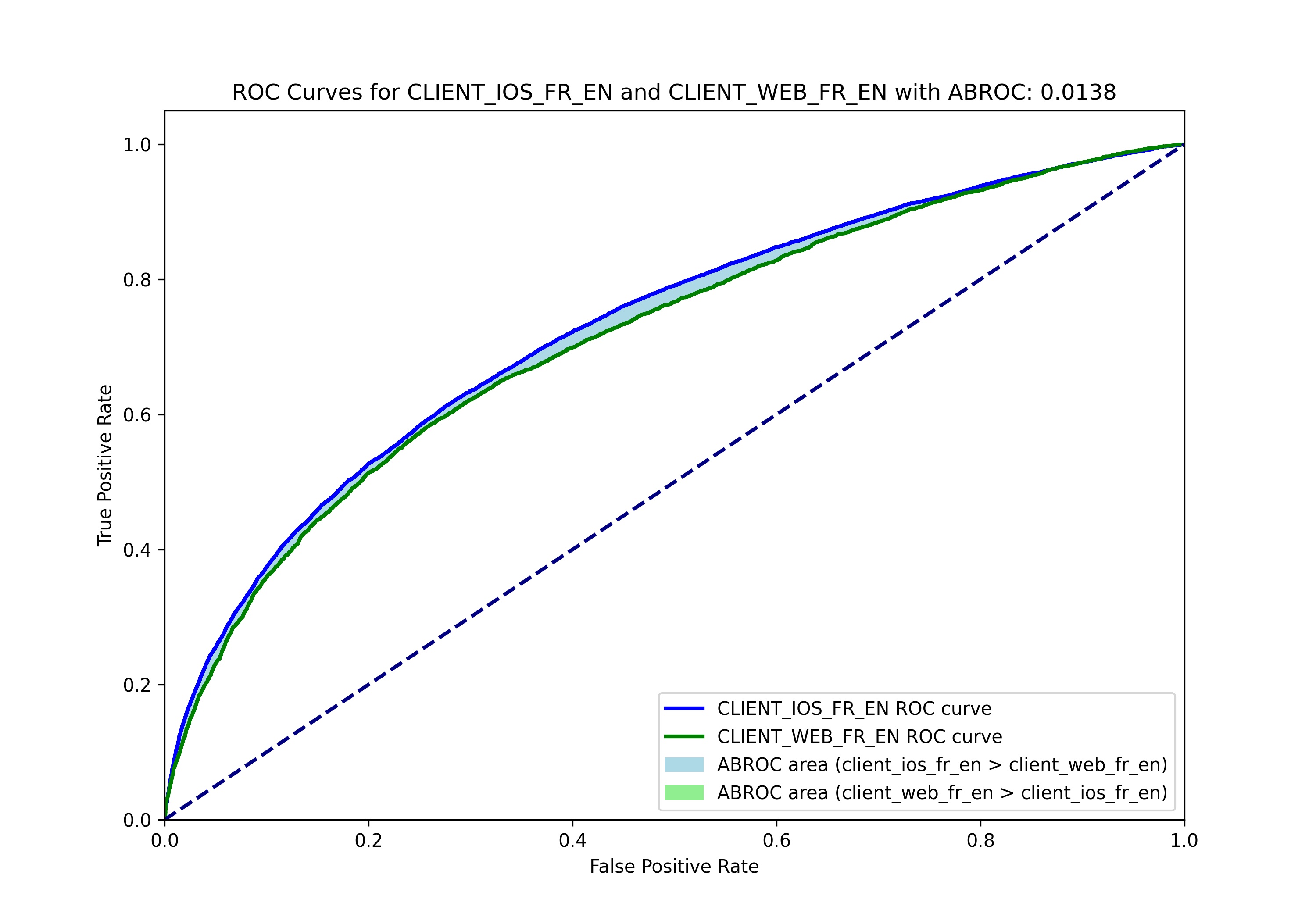}
    \caption{Fairness between ios and web in \texttt{fr\_en} track for Multi-task learning}
    \label{fig:client_ios_fr_en_VS_client_web_fr_en_ROC_ABROC}
\end{minipage}\hfill
\end{figure}

\clearpage
\textbf{All the graphs related to fairness on Country are listed below}
\begin{figure}[htbp]
\centering
\begin{minipage}[t]{0.45\textwidth}
    \centering
    \includegraphics[width=\linewidth]{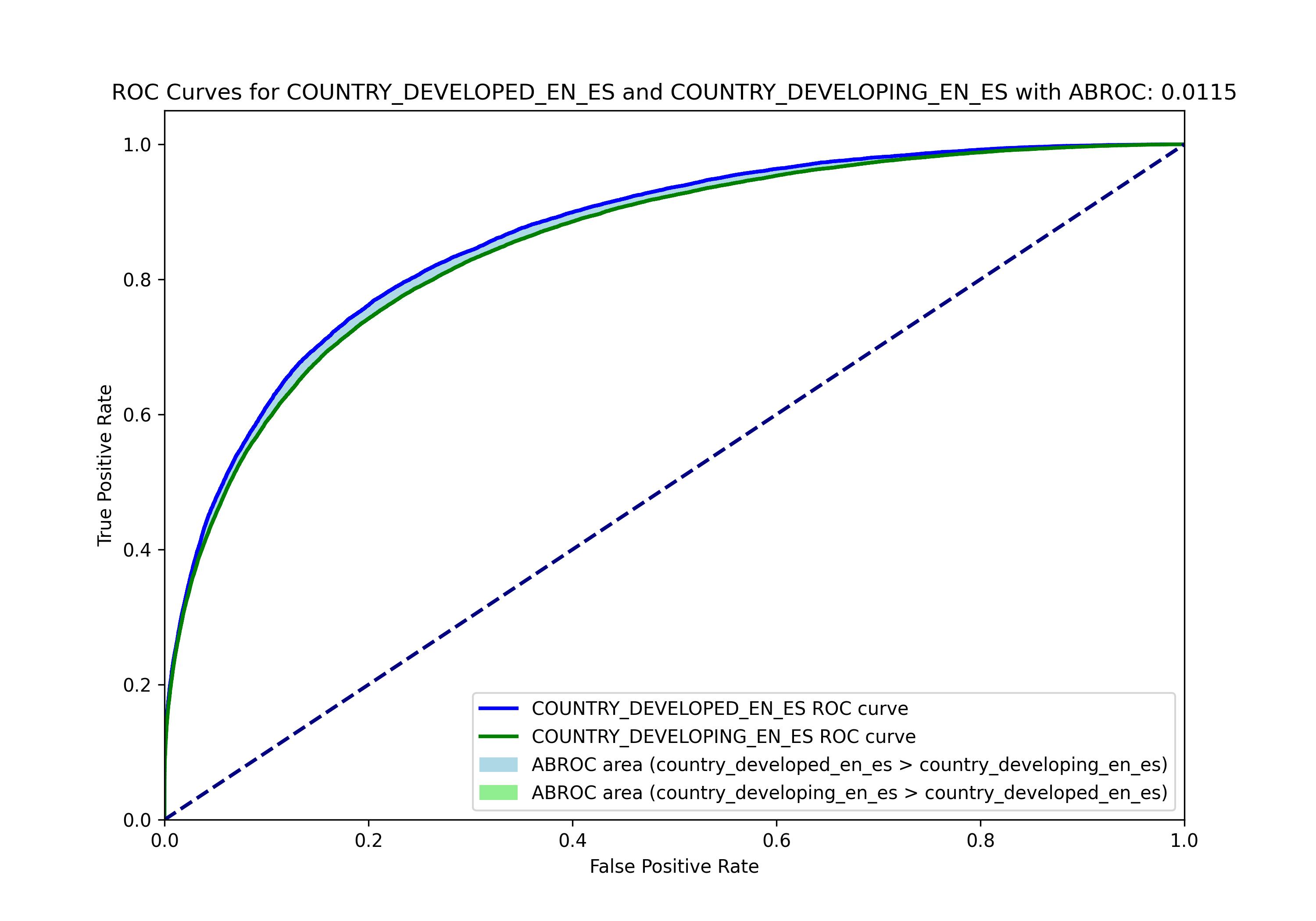}
    \caption{Fairness between developed and developing in \texttt{en\_es} track for GBDT}
    \label{fig:country_developed_en_es_VS_country_developing_en_es_ROC_ABROC}
\end{minipage}\hfill
\begin{minipage}[t]{0.45\textwidth}
    \centering
    \includegraphics[width=\linewidth]{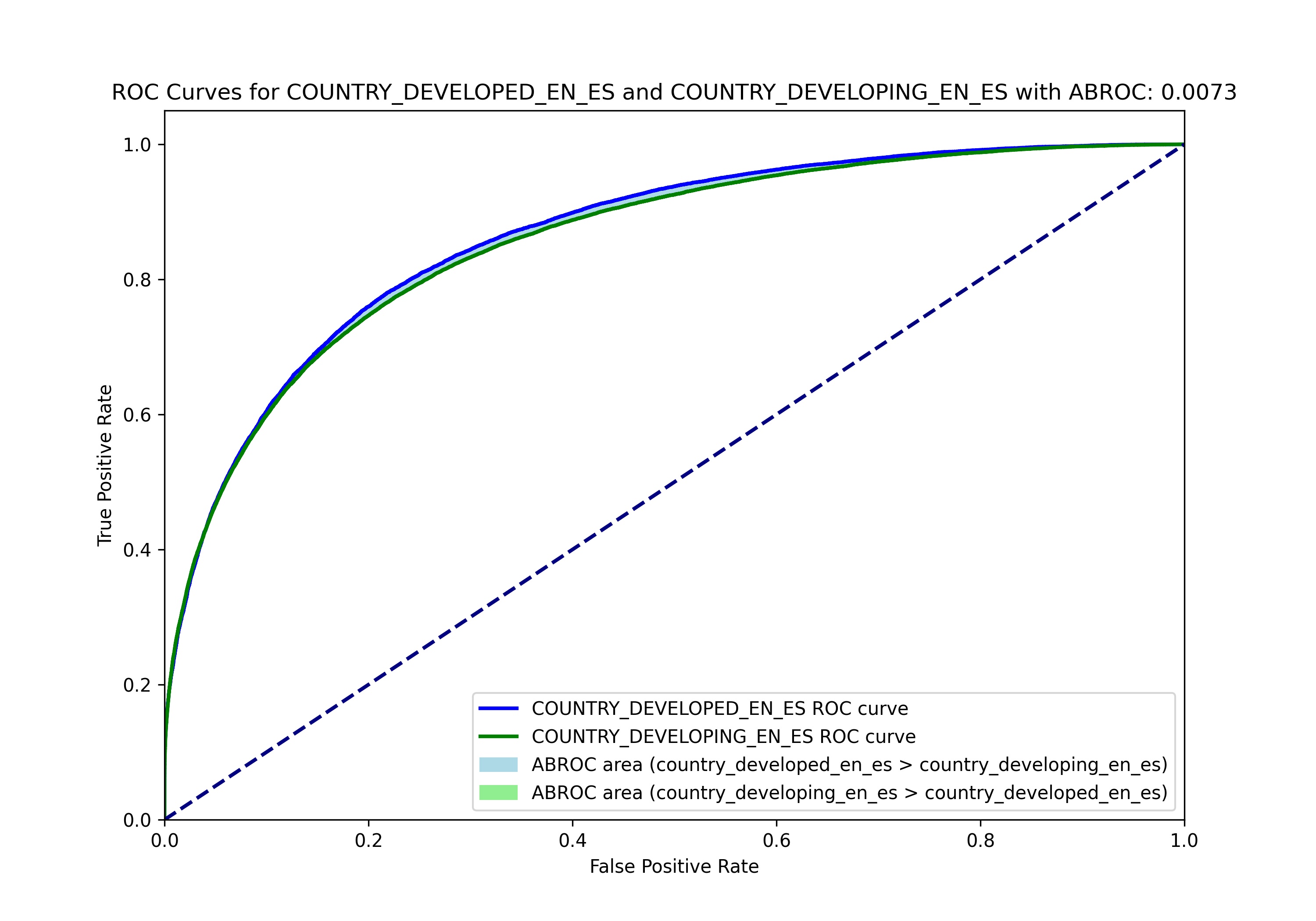}
    \caption{Fairness between developed and developing in \texttt{en\_es} track for Multi-task learning}
    \label{fig:country_developed_en_es_VS_country_developing_en_es_ROC_ABROC}
\end{minipage}\hfill
\end{figure}

\begin{figure}[htbp]
\centering
\begin{minipage}[t]{0.45\textwidth}
    \centering
    \includegraphics[width=\linewidth]{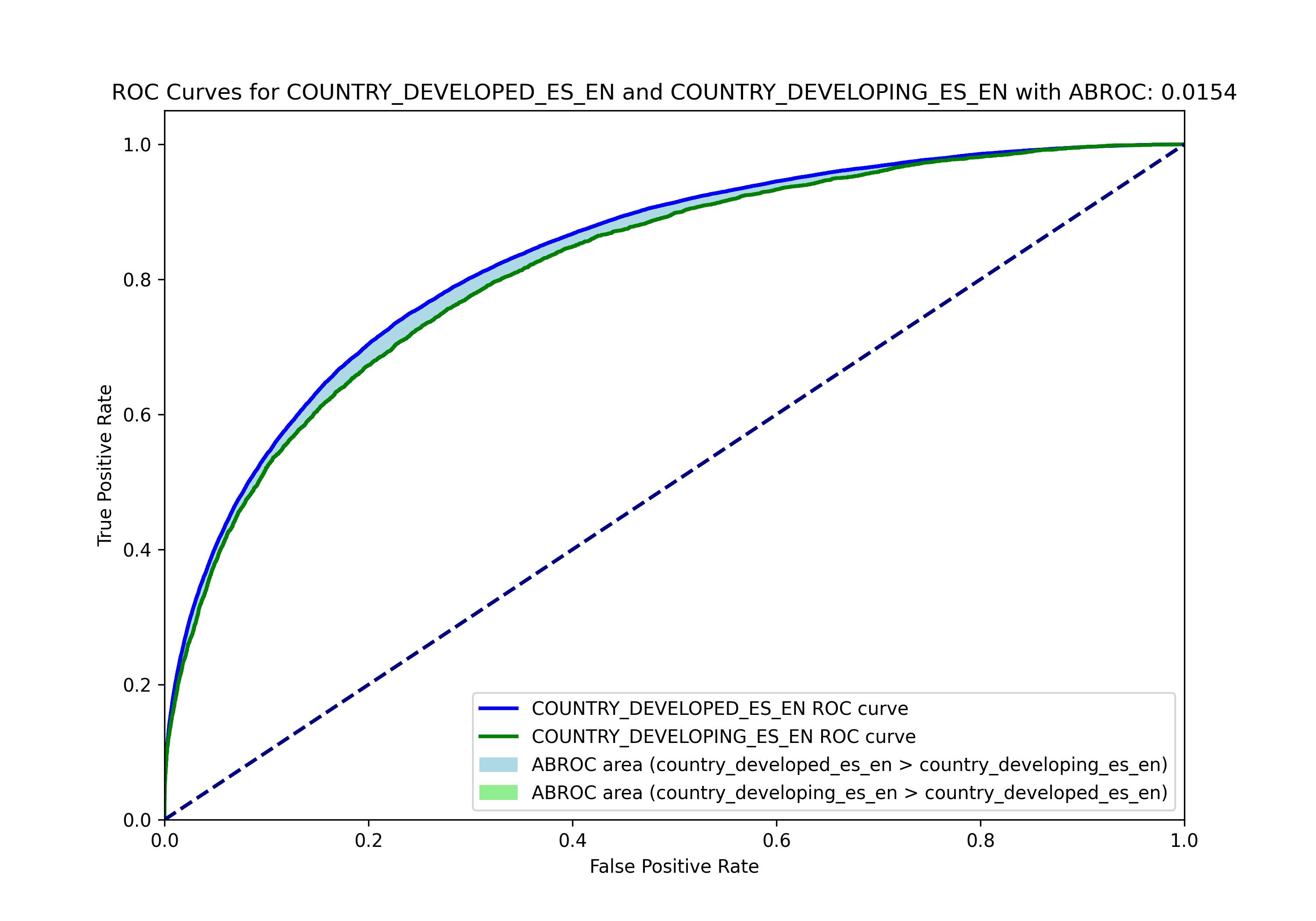}
    \caption{Fairness between developed and developing in \texttt{es\_en} track for GBDT}
    \label{fig:country_developed_es_en_VS_country_developing_es_en_ROC_ABROC}
\end{minipage}\hfill
\begin{minipage}[t]{0.45\textwidth}
    \centering
    \includegraphics[width=\linewidth]{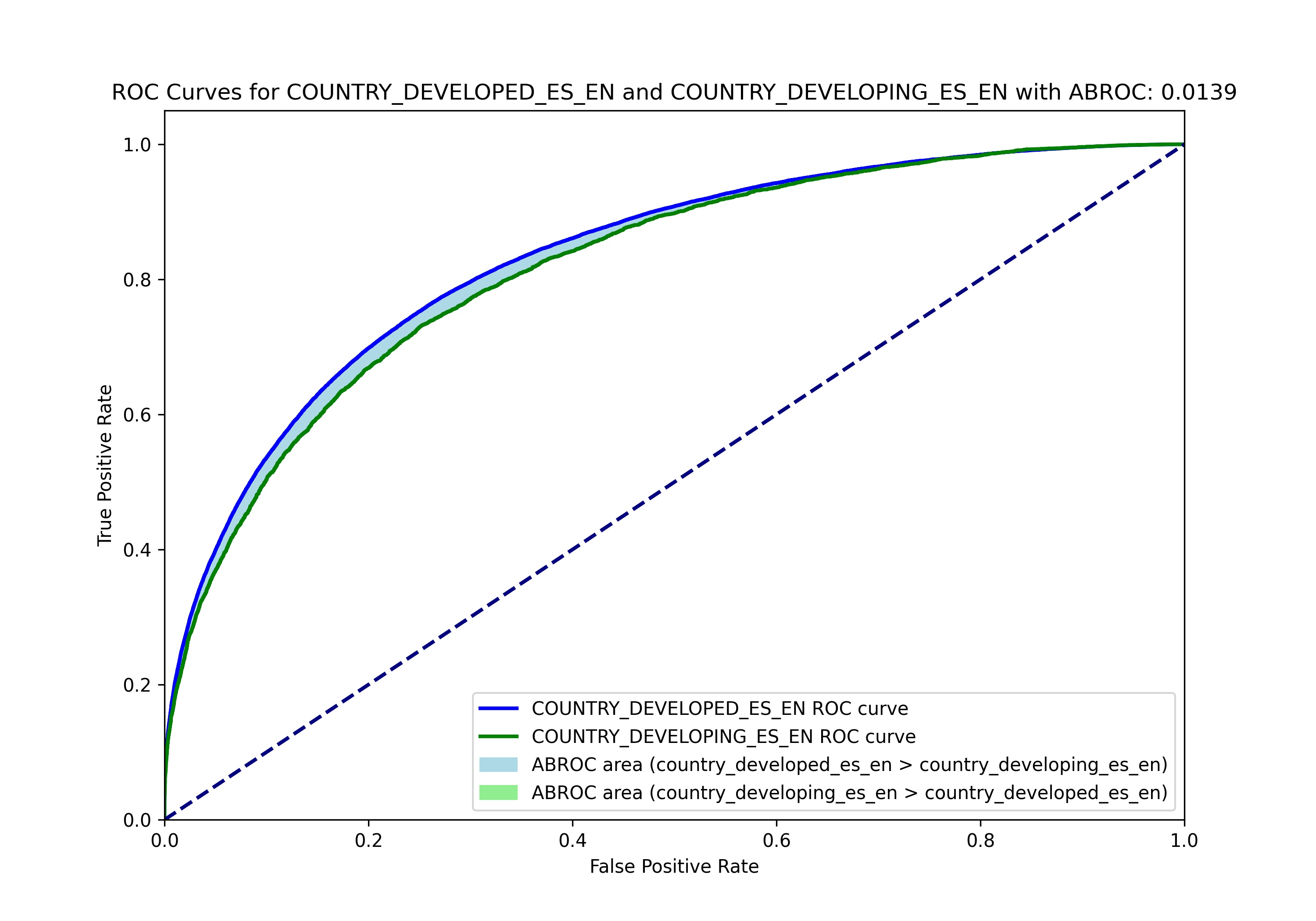}
    \caption{Fairness between developed and developing in \texttt{es\_en} track for Multi-task learning}
    \label{fig:country_developed_es_en_VS_country_developing_es_en_ROC_ABROC}
\end{minipage}\hfill
\end{figure}

\begin{figure}[htbp]
\centering
\begin{minipage}[t]{0.45\textwidth}
    \centering
    \includegraphics[width=\linewidth]{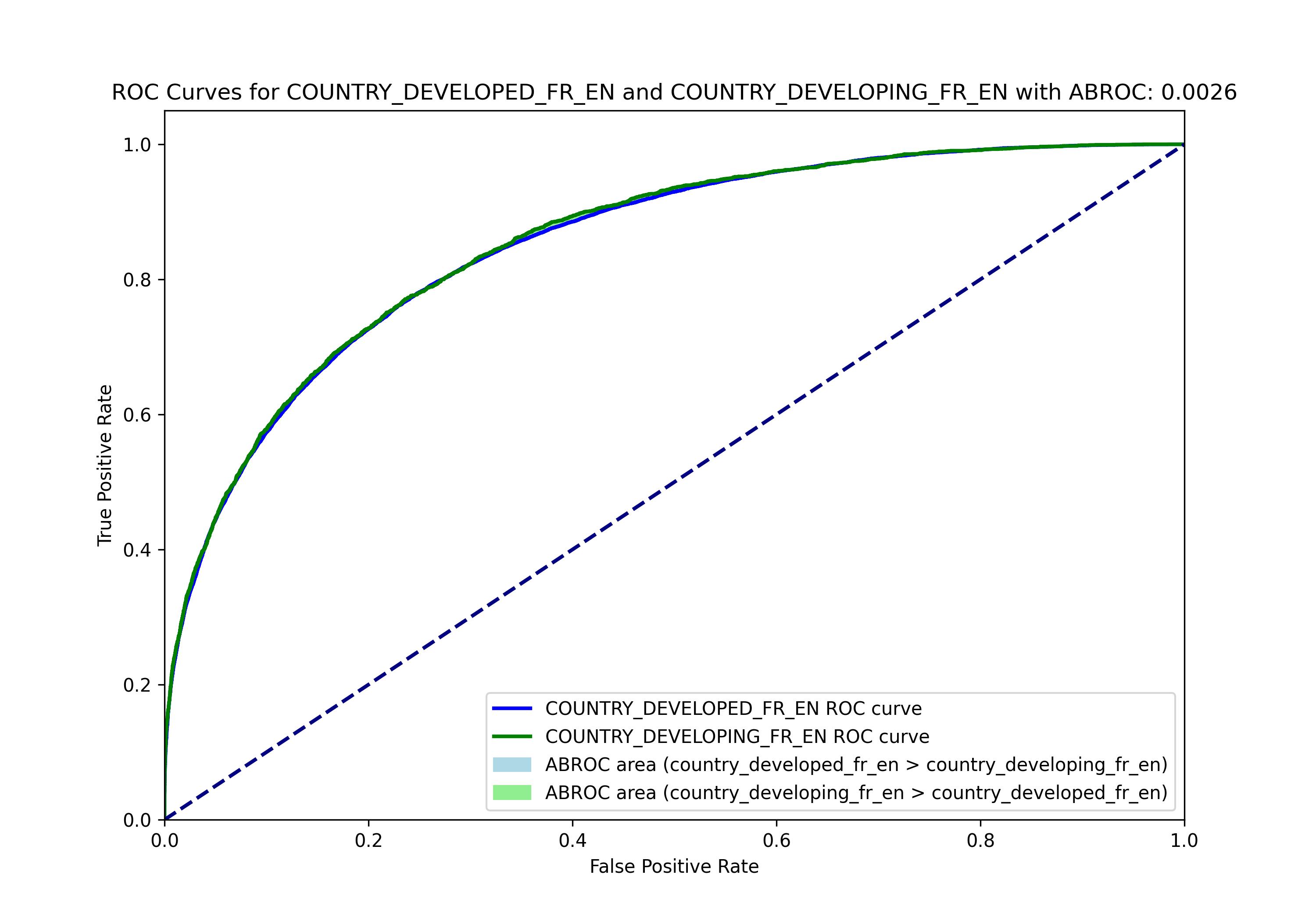}
    \caption{Fairness between developed and developing in \texttt{fr\_en} track for GBDT}
    \label{fig:country_developed_fr_en_VS_country_developing_fr_en_ROC_ABROC}
\end{minipage}\hfill
\begin{minipage}[t]{0.45\textwidth}
    \centering
    \includegraphics[width=\linewidth]{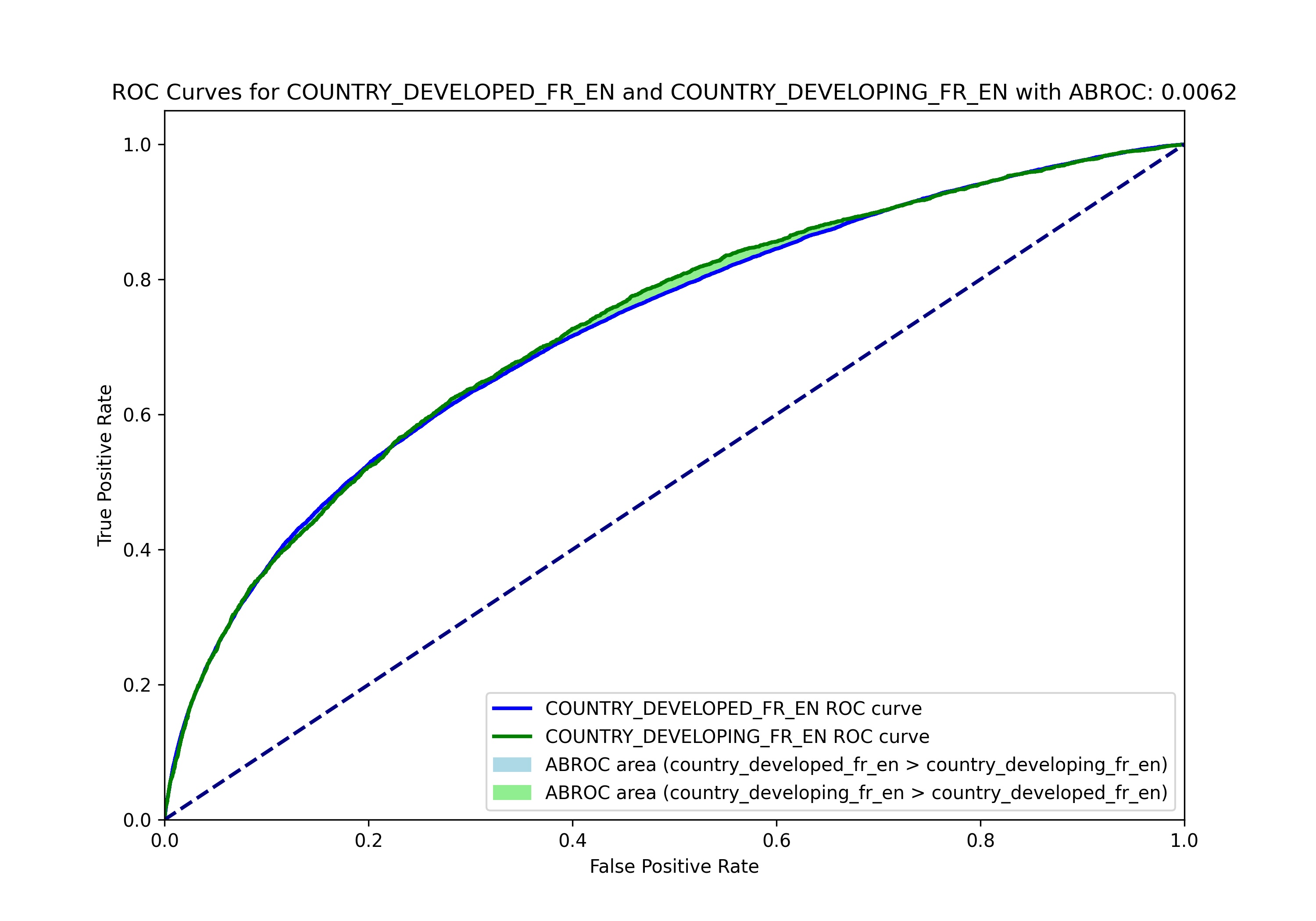}
    \caption{Fairness between developed and developing in \texttt{fr\_en} track for Multi-task learning}
    \label{fig:country_developed_fr_en_VS_country_developing_fr_en_ROC_ABROC}
\end{minipage}\hfill
\end{figure}
%%%%%%%%%%%%%%%%%%%%%%%%%%%%%%%%%%%%%%%%%%%%%%%%%%%%%%%%%%%%

\clearpage

\end{document}